\documentclass[wrr]{AGUTeX}

\usepackage{graphicx}
\usepackage{amsfonts,amsmath,amsthm}

\setkeys{Gin}{draft=false}

\authorrunninghead{F.~J. CORNATON}
\titlerunninghead{TRANSIENT WATER AGE DISTRIBUTIONS IN ENVIRONMENTAL FLOW SYSTEMS}
\authoraddr{F.~J. Cornaton,
Water Center for Latin America and the Caribbean, Tecnol\'{o}gico de Monterrey,
Eugenio Garza Sada 2501 Sur, 64849 Monterrey, N.L., M\'{e}xico.
(fabien.cornaton@itesm.mx)}

\begin{document}

\title{\textbf{Transient Water Age Distributions in Environmental Flow Systems: The Time-Marching Laplace Transform Solution Technique.}}

\authors{F.~J. Cornaton \altaffilmark{1}}
\altaffiltext{1}{Water Center for Latin America and the Caribbean, Tecnol\'{o}gico de Monterrey,
Monterrey, M\'{e}xico.}

\begin{abstract}
Environmental fluid circulations are very often characterized by analyzing the fate and behavior of natural and anthropogenic tracers. Among these tracers, age is taken as an ideal tracer which can yield interesting diagnoses, as for example the characterization of the mixing and renewal of water masses, of the fate and mixing of contaminants, or the calibration of hydro-dispersive parameters used by numerical models. Such diagnoses are of great interest in atmospheric and ocean circulation sciences, as well in surface and subsurface hydrology. The temporal evolution of groundwater age and its frequency distributions can display important changes as flow regimes vary due to natural change in climate and hydrologic conditions and/or human induced pressures on the resource to satisfy the water demand. Groundwater age being nowadays frequently used to investigate reservoir properties and recharge conditions, special attention needs to be put on the way this property is characterized, would it be using isotopic methods or mathematical modelling. Steady-state age frequency distributions can be modelled using standard numerical techniques, since the general balance equation describing age transport under steady-state flow conditions is exactly equivalent to a standard advection-dispersion equation. The time-dependent problem is however described by an extended transport operator that incorporates an additional coordinate for water age. The consequence is that numerical solutions can hardly be achieved, especially for real 3-D applications over large time periods of interest. A novel algorithm for solving the age distribution problem under time-varying flow regimes is presented and, for some specific configurations, extended to the problem of generalized component exposure time. The algorithm combines the Laplace Transform technique applied to the age (or exposure time) coordinate with standard time-marching schemes. The method is validated and illustrated using analytical and numerical solutions considering 1-D, 2-D and 3-D theoretical groundwater flow domains.
\end{abstract}

\begin{article}
%
\section{Introduction}\label{sec:intro}
A comprehensive analysis of flow processes and advective-dispersive-diffusive transport of
dissolved constituents in natural fluid circulations is an important challenge in Earth and environmental sciences. An approach that has become
very popular consists in using water age as an ideal tracer to tag fluid masses and estimate
associated timescales~(\cite{Deleersnijder10}). Such timescales can lead to very helpful
diagnoses that apply to interdisciplinary environmental studies, like in atmospheric and ocean circulation sciences or surface and subsurface hydrology.\protect\\
The characterization of water age characteristics in environmental flow systems is linked to the estimation of recharge patterns and variations, fluid flow dynamics over possibly various time scales, advective-dispersive/diffusive transport in heterogeneous circulations, and to some extend isotope geochemistry, fractionation, and interfacial reactive chemistry~(\cite{Glynn05,Ginn09}). In particular the characterization of groundwater age distributions is of prime interest as it can act as an indicator of aquifer contamination and vulnerability, explain the modes of recharge of aquiferous reservoirs, or identify the diffusive exchanges between mobile and immobile flow regions. The literature offers a wide range of studies focusing on various aspects of water
age and its conceptual representations. However because different mathematical and conceptual foundations are used, a standardization of the results is still difficult to be achieved.\protect\\
Age is commonly inferred from concentrations of single
isotopes/tracers~(for exhaustive reviews of groundwater age dating methods e.g. see~\cite{Clarke97,Kazemi06}), or from multiple isotope/tracer concentrations that together represent different components of age~(e.g. see~\cite{Corcho07,Troldborg08,Larocque09,Lavastre10}). This concept is however very misleading because concentration-based ages are simply apparent ages that do not a priori relate to the mean age of a water sample~(\cite{Sanford11}). Moreover, the use of environmental and anthropogenic tracers to make inferences on water age goes together with the use of lumped-parameter models that assume specific age distributions regardless the complexity of the flow domain. \protect\\
A relatively recent representation of age and its distributions is the one that relies on the use of distributed flow models. Accounting for the relevant transport characteristics affecting age distributions however requires the use of adequate governing transport equations. Regarding this point, new advances have permitted to consolidate the unification of the foundations on age fate by means of a physically-based derivation of a transport equation for mean age~(\cite{Goode96}), percentile of age and moments of age~(\cite{Varni98}), and later on, after the early works of~\cite{Campana87}, of equations describing the complete age frequency distribution in the context of ocean circulations modelling~(\cite{Deleersnijder01,Delhez02}) and subsurface hydrology~(\cite{Ginn99,Ginn00a,Ginn00b,Cornaton03,Cornaton06a,Cornaton06b,Cornaton07a,Ginn07,Woolfenden09}).\protect\\
Putting aside the way age is inferred (or modelled), the transient nature of flow systems and its implication on the transient nature of age distributions have never been successfully addressed, particularly for large space- and time-scales and real-site applications. This aspect is however of high importance since the actual situations at which age properties are measured are a result of past-to-present perturbations~(\cite{Ginn09}), at short and large time-scales.~\cite{Bethke08} correctly pointed out that changing the field of groundwater age dating is a means for thinking about groundwater age in a new way. We can additionally argue that the way age is modelled is also to be refined. In particular the time-dependency of age distributions play a fundamental role in the interpretation of age and tracer data. Transient groundwater age distributions are for example a result of natural transient hydrologic conditions and of human-induced modifications on the water cycles by artificial withdraw and recharge, but they can also originate from more specific phenomena as the temporal dependency of fluid flow to fluid density and viscosity (e.g. in thermohaline problems). When large time-scales are considered, the hydrodynamic conditions of a system are changing in time, in relation to the evolutions of geomorphology, system internal structure, climate change and thus recharge conditions. Any modification induces changes in the age distributions.\protect\\
This article presents the outcomes of research works on transient age distributions and their numerical solutions. We present a novel numerical formulation that is able to handle this problem in 3-D discretized domains of arbitrary complexity, and that applies to the most relevant numerical integration techniques such as the finite element, finite volume, and finite differences methods. The method is also particularly suited to fulfill the needs related to the modeling of hydrologic evolutions over very large time periods. In the first section we summarize the mathematical foundations of transient age and constituent exposure time density equations. In the second section we make a short review of existing numerical integration schemes and present a novel numerical formulation. The characteristics of the proposed algorithms are analyzed and discussed, and eventually validated and illustrated in the last section using analytical and numerical solutions.
%
\section{Mathematical Formulations}\label{sec:basics}
This section summarizes the theoretical basics for modeling water age and constituent exposure time distributions under transient conditions of fluid flow.

\subsection{Equations for the Distribution of Constituent Exposure Time and Water age}\label{sec:agepdf}
The time spent by constituent molecules undergoing transport in a given flow domain (e.g. water, solutes, suspended
colloids) can be defined as an independent dimension over which elemental masses are distributed, just as they are distributed over a spatial dimension (see~\cite{Ginn99,Delhez02}). Following~\cite{Ginn99}, the exposure time is defined as this independent dimension, and it is convenient to define the space over which such density distributions evolve as the $n+1$ dimensional space, that is elaborated by augmenting the $n$ physical dimensions with one real independent dimension, say $\tau$, that represents an exposure time used by constituent molecules in a given reservoir. This additional dimension is a continuous measure orthogonal to the remaining coordinates. Letting $\emph{\textbf{x}} = (x, y, z)$ be the physical coordinates for the case $n = 3$, and $t$ be the clock-time, a point $(\emph{\textbf{x}}, t)$ in the physical distance and time dimensions lies only in the 4-D space $\mathbb{R}^{n} \times \mathbb{R}^{1}$, while a point $(\emph{\textbf{x}}, t, \tau)$ lies in the augmented 5-D space $\mathbb{R}^{n+1} \times \mathbb{R}^{1}$. \protect\\
The equation governing the distribution of constituent $\alpha$ exposure time density $\rho_{\alpha} = \rho_{\alpha}(\textit{\textbf{x}},t,\tau)$ within a flow domain $\Omega$ is given by~\cite{Ginn99}:
\begin{equation}\label{eq:5Ddensity}
\frac{\partial \rho_{\alpha}}{\partial t} + \nabla \cdot \textbf{j}_{\alpha} + \frac{\partial \upsilon_{\alpha} \rho_{\alpha}}{\partial \tau} = r_{\alpha}
\end{equation}
in which $\textbf{j}_{\alpha} = \textbf{j}_{\alpha,A} + \textbf{j}_{\alpha,D}$ is the total mass density flux that accounts for advective flux $\textbf{j}_{\alpha,A} = \textbf{v} \rho_{\alpha}$ and diffusive
and pore-scaled dispersive flux $\textbf{j}_{\alpha,D}$. The physical velocity field $\textbf{v}(\emph{\textbf{x}},t)$ derives from a pre-solution of an appropriate fluid flow equation. We use the Fickian constitutive theory for diffusion and dispersion in order to express $\textbf{j}_{\alpha,D}$ as a
potential in mass density, $\textbf{j}_{\alpha,D} = - \textbf{D} \nabla \rho_w$, with $\textbf{D}(\emph{\textbf{x}},t)$ being a conventional dispersion-diffusion tensor. The total mass density flux is then expressed as $\textbf{j}_{\alpha} = \textbf{v} \rho_{\alpha} - \textbf{D} \nabla \rho_w$. The exposure-time velocity $\upsilon_{\alpha}$ is the rate of change of a material point location on the exposure-time direction $\tau$. The reaction term $r_{\alpha}$ represents the material mass density rate of transformation.\protect\\
\begin{figure}[h]
\noindent \includegraphics[width=0.5\textwidth]{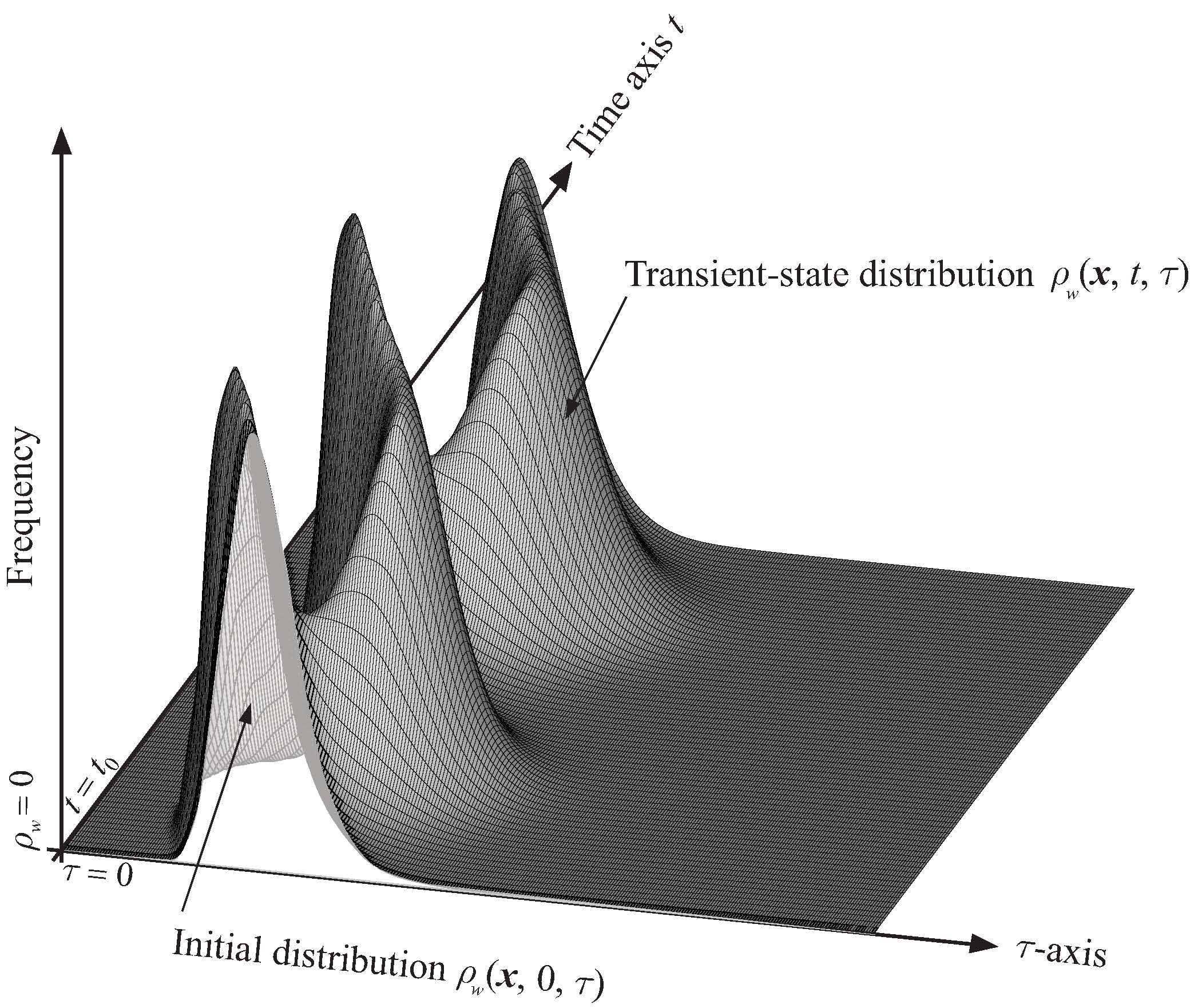}
\caption{Illustration of a transient distribution of water age as deduced from an oscillating uniform one-dimensional velocity field.\label{fig:fig1}}
\end{figure}

For the case of water age ($\alpha = w$), the density $\rho_w = \rho_w(\textit{\textbf{x}},t,\tau)$ is the space- and time-dependent bulk-volumetric mass density of water distributed over age $\tau$, and the exposure time velocity $\upsilon_w$ in the $\tau$-direction is simply 1. Consequently, the equation governing water age density simplifies in
\begin{equation}\label{eq:5Dage0}
\frac{\partial \rho_w}{\partial t} + \nabla \cdot \textbf{j}_w + \frac{\partial \rho_w}{\partial \tau} = r_w
\end{equation}
With equation~(\ref{eq:5Dage0}), age distributes by the effect of the advection at a unit velocity of water mass density along the age axis, and the age variation is expressed by the term $\frac{\partial \rho_w}{\partial \tau}$. Age mixing is classically ruled by the variations of fluid velocity $\textbf{v}$ and by the processes of dispersive/diffusive mixing represented by the $\textbf{D}$-tensor. A theoretical density distribution of water age density distribution $\rho_w(\emph{\textbf{x}},t,\tau)$ at a given position $\emph{\textbf{x}}$ is displayed in Figure~\ref{fig:fig1}. The distribution undergoes a temporal evolution from an initial situation that may correspond to an equilibrated (or steady) state.

\subsection{The Case of Groundwater Age}\label{sec:gwagepdf}
The specific context of subsurface flow implies fluid motion through heterogeneous porous media of porosity (or mobile water content) $\theta(\emph{\textbf{x}}, t)$. The flow equation yielding the distribution of fluid flux $\textbf{q}(\emph{\textbf{x}},t) = \theta(\emph{\textbf{x}},t) \textbf{v}(\emph{\textbf{x}},t)$ can be written using the modified form of Richards' equation, which describes the
three-dimensional transient groundwater flow in a variably-saturated porous matrix:
\begin{equation}\label{eq:flowequation}
\frac{\partial \theta}{\partial t} + \nabla \cdot \textbf{q} = q_\textrm{I} - q_\textrm{O}
\end{equation}
in which the fluid flux vector is $\textbf{q} = - k_r \textbf{K} \nabla H$ with $H(\textit{\textbf{x}},t) = h(\textit{\textbf{x}},t) + z$ being the hydraulic head and $h(\textit{\textbf{x}},t)$ the pressure head, and where \textbf{K} is the saturated tensor of hydraulic conductivity, $k_r = k_r(S_w)$ is the relative permeability of the medium with respect to the degree of water saturation $S_w = \frac{\theta}{\theta_s}$, with $\theta_s$ being the saturated water content. The terms $q_\textrm{I}$ and $q_\textrm{O}$ are fluid source and sink terms, respectively.\protect\\

Using the definition of the bulk-volumetric water mass density $\rho_w(\emph{\textbf{x}},t,\tau) = \theta(\emph{\textbf{x}},t) g(\emph{\textbf{x}},t,\tau)$, with $g(\emph{\textbf{x}},t,\tau)$ being the frequency distribution of groundwater ages, and formulating the dispersive component of mass flux as $- \theta \textbf{D} \nabla g$, equation~(\ref{eq:5Dage0}) transforms into
\begin{equation}\label{eq:5Dage1}
\frac{\partial \theta g}{\partial t} = - \nabla \cdot \left ( \textbf{q} g - \theta \textbf{D} \nabla g \right ) - \frac{\partial \theta g}{\partial \tau} + r
\end{equation}
in which the term $r = q_\textrm{I} \delta(\tau) - q_\textrm{O} g$ accounts for the divergence of fluid flux in equation~(\ref{eq:flowequation}). Equation~(\ref{eq:5Dage1}) is subject to the following initial and boundary conditions:
\begin{subequations}\label{eq:5Dage_ICBC}
\begin{equation}\label{eq:ica}
g(\emph{\textbf{x}},0,\tau) = f_0(\emph{\textbf{x}},\tau) \quad\quad \text{in} \quad \Omega
\end{equation}
\begin{equation}\label{eq:5Dage_BC1}
g(\emph{\textbf{x}},t,\tau) = \delta(\tau) \quad\quad \text{on} \quad \Gamma_{-,1}
\end{equation}
\begin{equation}\label{eq:5Dage_BC2}
-\theta \textbf{D} \nabla g \cdot \textit{\textbf{n}} = f_2(t,\tau) \quad\quad \text{on} \quad \Gamma_{-,2}
\end{equation}
\begin{equation}\label{eq:5Dage_BC3}
\textbf{J} \cdot \textit{\textbf{n}} = \textbf{q} \delta(\tau) \cdot \textit{\textbf{n}} \quad\quad \text{on} \quad \Gamma_{-,3}
\end{equation}
\begin{equation}\label{eq:5Dage_BC0}
\textbf{J} \cdot \textit{\textbf{n}} = 0 \quad\quad \text{on} \quad \Gamma_0
\end{equation}
\end{subequations}
where $f_0(\emph{\textbf{x}},\tau)$ is a given function representing the (initial) age distribution function at time $t = 0$, \textit{\textbf{n}} is a normal outward
unit vector, and where the total age flux is defined by $\textbf{J} = \textbf{q} g - \theta \textbf{D} \nabla g$. The boundary portions $\Gamma_0$ and $\Gamma_-$ denote the no-flow and inflowing boundaries of $\Omega$, respectively, and the indices 1, 2 and 3 refer to boundary portions along which Dirichlet, Neumann and Cauchy age conditions are prescribed, respectively. The zero-age boundary condition is mathematically represented by the delta distribution $\delta(\tau)$, which ensures a pure unit and instantaneous pulse
input along the recharge boundary portions $\Gamma_-$. Since the age-coordinate is totally independent on the time-coordinate, this condition is simply assigned in a natural way. Even if both the Dirichlet and Cauchy conditions~(\ref{eq:5Dage_BC1}) and~(\ref{eq:5Dage_BC3}) can be combined, the Cauchy condition appears to be the one that is the most meaningful from a physical point of view, because it prevents water molecules from moving upstream and exiting the system by the inlet limits (it is homogeneous for $\tau > 0$). The Dirichlet-type condition is also homogeneous for $\tau > 0$; however it does not prevent the backward movement of water particles at inflowing limits when dispersion is significant. Moreover, the use of the Dirichlet-type condition can yield unphysical results when some parts of the inlet boundaries become inactive (i.e. during phases when recharge at the boundaries is not present). The Cauchy condition is active as long as inflow rates are non-zero ($\textbf{q} \cdot \textit{\textbf{n}} \neq 0$ in equation~(\ref{eq:5Dage_BC3})), and consequently it can be handled naturally without any special care, while the Dirichlet condition implies the controlled action of deactivating the condition where and when inflow limits stop producing, and of reactivating it as soon as these limits start to produce water again. The Neumann-type condition~(\ref{eq:5Dage_BC2}) can also be used at particular boundary portions where the production of age mass is driven by diffusion only. The function $f_2$ represents a purely diffusive production of age mass along $\Gamma_2$ boundary portions.\protect\\
Particular cases for which the internal production of water is associated to more specific, not necessary equal to zero distributions of age $\tau_I$ will require the volumetric fluid source intensity $q_\textrm{I}$ to be associated to an age distribution $g_I(\emph{\textbf{x}}, t, \tau) = \delta(\tau - \tau_I(t))$, being itself potentially time-dependent, such that the age mass source will be distributed as $q_\textrm{I}(\emph{\textbf{x}}, t) \delta(\tau - \tau_I(t))$.\protect\\

The equation describing the evolution of mean age is obtained by taking the first age moment of equation~(\ref{eq:5Dage1}), since the mean age $a(\emph{\textbf{x}},t)$ of water molecules in a REV is obtained by taking the first age moment of the age density distribution function $g(\emph{\textbf{x}},t,\tau)$:
\begin{equation}\label{eq:meanage}
a(\emph{\textbf{x}},t) = \int_0^{\infty} {\tau g(\emph{\textbf{x}},t,\tau) d\tau}
\end{equation}
Taking the first order age moment of equations~(\ref{eq:5Dage1}) and~(\ref{eq:5Dage_ICBC}) yields the following boundary value problem for mean age:
\begin{subequations}\label{eq:MA BVP}
\begin{equation}\label{eq:MA ADE}
\frac{\partial \theta a}{\partial t} = - \nabla \cdot \left ( \textbf{q} a - \theta \textbf{D} \nabla a \right ) - q_\textrm{O} a + \theta
\end{equation}
\begin{equation}\label{eq:MAic}
a(\emph{\textbf{x}},0) = a_0(\emph{\textbf{x}}) \quad\quad \text{in} \quad \Omega
\end{equation}
\begin{equation}\label{eq:MA_BC1}
a(\emph{\textbf{x}},t) = 0 \quad\quad \text{on} \quad \Gamma_{-,1}
\end{equation}
\begin{equation}\label{eq:MA_BC2}
-\theta \textbf{D} \nabla a \cdot \textit{\textbf{n}} = f_2(t) \quad\quad \text{on} \quad \Gamma_{-,2}
\end{equation}
\begin{equation}\label{eq:MA_BC3}
\textbf{J}_a \cdot \textit{\textbf{n}} = 0 \quad\quad \text{on} \quad \Gamma_{-,3} \, \cup \, \Gamma_0
\end{equation}
\end{subequations}
where mean age flux is defined as $\textbf{J}_a = \textbf{q} a - \theta \textbf{D} \nabla a$. Eq.~(\ref{eq:MA ADE}) corresponds to the equation derived by~(\cite{Goode96}), in which mean age is continuously generated
during water flow since porosity is acting as a source term. This source term
indicates that water is aging one unit per unit time, in
average. Equation~(\ref{eq:MA ADE}) is exactly equivalent to a standard advection-dispersion equation, and can be solved by any standard code that can produce advective-dispersive solute transport solutions. The only particularity resides in a mass source term that needs to be distributed as porosity is. Higher-order moment equations can also be derived from equation~(\ref{eq:5Dage1}) in analogy to the development yielding equation~(\ref{eq:MA ADE}) to eventually recover the moment equations derived by~\cite{Varni98} or~\cite{Delhez02}.
%
\section{Numerical Solutions of the Transient Age Distribution Equation}\label{sec:numerical}
In this section we first make a short review of existing numerical schemes yielding numerical solution of the transient age distribution problem, and discuss their limitations. A novel integration scheme is then presented and analyzed in terms of its adequacy for handling problems of arbitrary complexity and configuration.\protect\\

Steady-state models of flow regimes involve steady-state age distributions, with which important simplifications are possible since the
steady age distribution satisfies the time-translation invariance $g(\emph{\textbf{x}}, t, t - \tau) = g(\emph{\textbf{x}}, 0, -\tau), \quad \forall t \in [0,\ldots,\infty[$ (\cite{Haine08}). This property means that steady distributions only depend on the age coordinate $\tau$. As a consequence, the steady-state form of the age distribution equation~(\ref{eq:5Dage1}) is exactly equivalent to an advection-dispersion equation of transient nature, and can be solved using standard advection-dispersion solvers (see~\cite{Cornaton03},~\cite{Cornaton06a},~\cite{Cornaton06b}):
\begin{equation}\label{eq:4Dage}
\frac{\partial \theta g}{\partial \tau} = - \nabla \cdot \left ( \textbf{q} g - \theta \textbf{D} \nabla g \right ) + r
\end{equation}
The only technical difficulty for solving equation~(\ref{eq:4Dage}) is related to the type of boundary condition required to properly describe the zero-age condition. This zero-age condition is assumed to be Dirac along inflow boundaries, and is classically modelled by means of a Dirichlet-type condition $g(\emph{\textbf{x}}, \tau) = \delta(\tau)$ and/or a Cauchy-type mass flux condition $\textbf{J} \cdot \emph{\textbf{n}} = \textbf{q}(\emph{\textbf{x}})\delta(\tau) \cdot \emph{\textbf{n}}$. In order to properly handle these conditions,~\cite{Cornaton03} proposed to make use of Laplace transforms (since the function $\delta(\tau)$ transforms into 1) and of the LTG formalism (\cite{Sudicky89}) to obtain finite element solutions.\protect\\

Unsteady hydrodynamic conditions involve unsteady age distributions with which the time-translation invariance inherent in equation~(\ref{eq:4Dage}) is not valid anymore, such that age is now distributed along two independent temporal dimensions: clock-time $t$ and age $\tau$. This fact means that an age distribution for each clock-time and field point has to be considered and, as a consequence, the 5-D transient age distribution equation presents severe technical difficulties in its numerical resolution in terms of computational burden and, above all, in terms of memory usage. In fact, the additional dimension induced by the age dimension requires the storage of a complete distribution $g(\tau)$ at any computational point of the system, and for all considered simulation times.\protect\\
To figure out the enormous computational requirements involved by the solution of the transient age distribution equation, assume that we wish to represent
$g(\emph{\textbf{x}}, t, \tau)$ as a discrete matrix that is discretized
with $N_{\emph{\textbf{x}}}$ points (or cells) over field points $\emph{\textbf{x}}$, $N_t$ points over
clock-time $t$, and $N_{\tau}$ points over age $\tau$. On the one hand, the storage requirements do not
depend on the discretization method and are directly proportional to $N_{\emph{\textbf{x}}} N_t N_{\tau}$. On the other hand, the computer time required to compute the $g-$matrix does depend on the choice of the method. Modern numerical models can generally handle up to
$N_{\emph{\textbf{x}}} = \textrm{O}(10^{6-7})$ and $N_t \approx N_{\tau} = \textrm{O}(10^{4-7})$. The resulting $g-$matrix therefore has
$\textrm{O}(10^{14-21})$ elements and is obviously impossible to store. \protect\\

\cite{Ginn99} proposed to treat the first-order term $\frac{\partial g}{\partial \tau}$
in analogy to a physical advective process, thus grouping this
term with the advection term $\nabla \cdot \textbf{q} g$. That is, the first-order terms of equation~(\ref{eq:5Dage1}) are taken together as $\nabla \cdot \textbf{q}_a g$, where $\textbf{q}_a = \{\textbf{q}, 1\}$ is the velocity vector augmented by the age dimension, which presents the important property that its divergence in the $(n+1)$-dimensions space is given by the divergence in the standard Cartesian space of dimension $n$ (\cite{Ginn99}). A direct consequence of this formulation is that the dispersion tensor $\textbf{D}$ also needs to be augmented with another dimension by adding a column and a row of zero entries ($\textbf{D}_a = \{\textbf{D}, 0\}$). Using the augmented velocity vector and dispersion tensor, equation~(\ref{eq:5Dage1}) becomes equivalent to an advection-dispersion equation, $\frac{\partial \theta g}{\partial t} = - \nabla \cdot \left ( \textbf{q}_a g - \theta \textbf{D}_a \nabla g \right ) + r$, which can be solved for $n < 3$ using conventional solute transport codes. The obvious restriction of such a treatment of the age coordinate is that the physical dimension $n$ cannot be bigger than 2, that is, at best a 3-D Cartesian domain can be used to solve for equation~(\ref{eq:5Dage1}) in a 2-D spatial representation of a reservoir. Another difficulty is related to the need of a preliminary (optimal) definition for the size and the discretization of the physical mesh representing the age dimension. Since the variations in time of the ages are not known prior to calculation, a large enough grid has to be designed in order to capture the increase in age in time and a uniform discretization in the age direction for each physical mesh point needs to be defined (i.e. a uniform grid in the age direction). These aspects are quite restrictive and no ideal solution seems to exist. The technique is useful to provide validation examples for other approaches, but is obviously not practical for real-site applications.\protect\\

In the context of the modeling of oceanic circulations and surface water bodies,~\cite{Delhez02} also proposed a numerical solution approach, but their scheme has never proved to be efficient to handle large domains and time-scales.
As already discussed in the previous section, the term $\frac{\partial \theta g}{\partial \tau}$ in equation~(\ref{eq:5Dage1}) can be considered as an advection term towards larger ages and, hence, discretized with the techniques usually applied to advection terms. Equation~(\ref{eq:5Dage1}) can therefore be solved as an evolution equation in the 5-D space $\mathbb{R}^{n+1} \times \mathbb{R}^{1}$, when using an operator splitting technique to handle separately the advection terms in the physical space and the age space (\cite{Delhez02}). When finite element, finite volume, or finite differences techniques are used, such a resolution involves not only the discretization of equation~(\ref{eq:5Dage1}) on a usual 3-D mesh or grid but also in the age dimension by splitting the age domain $\tau \in [0, \tau_{max}]$ into an appropriate number of age classes from minimum age 0 to a maximum age $\tau_{max}$.~\cite{Delhez02} propose to avoid the discretization of the ageing term and associated errors by introducing the change of variables $(t, \tau) \rightarrow (\hat t = t, \hat \tau = t - \tau)$, which yields $\frac{\partial \theta g}{\partial t} + \frac{\partial \theta g}{\partial \tau} + \cdots \rightarrow \frac{\partial \theta g}{\partial \hat t} + \cdots$, and thus removes the differential terms in the $\tau-$direction. The resolution of equation~(\ref{eq:5Dage1}) finally corresponds to solving a number of classical uncoupled advection-dispersion equations corresponding to the different considered discrete age classes. As opposed to the mesh-based method of~\cite{Ginn99}, the operator splitting technique of~\cite{Delhez02} has the advantage of being applicable to 3-D spaces, but it presents the disadvantage that a fixed number of age classes and a maximum allowed age have to be preliminary defined. The number of age classes can only be limited due to obvious memory storage limits, and this method may not operate with cases for which time-frames bigger than the maximum age present in the system have to be considered.

\subsection{The Time-Marching Laplace Transform Solution Technique (TMLT)}\label{sec:TMLTScheme}
In this section, a novel integration scheme is presented. The main aspect of the scheme is that a reduction in dimension
is achieved by mathematical transformation of the age dimension. The method combines the formalism of the LTG technique (\cite{Sudicky89}) and the techniques of time-marching classically used with mass transport solutions. The basic principle relies on the fact that age $\tau$ and clock-time $t$ are totally independent quantities, and that the rate of ageing does not depend on time $t$, since it is always one per unit time. In order to be relieved from the problem induced by the existence of an advection term in the age coordinate (see the two previous sections), and for the sake of reducing the 5-dimensional system to a 4-dimensional system, a linear integral transform is applied to the age dimension $\tau$. The function $g(\emph{\textbf{x}}, t, \tau)$ being locally integrable on $[0, \infty)$, we can make use of an unilateral Laplace transform since the necessary condition for its existence is fulfilled.\protect\\

When a Laplace transform is applied to the 5-D transient age distribution equation~(\ref{eq:5Dage1}) in the $\tau-$dimension, one obtains the following transformed transient age equation:
\begin{equation}\label{eq:TMLT_Eqn}
\frac{\partial \theta \hat g}{\partial t} = - \nabla \cdot \left ( \textbf{q} \hat g - \theta \textbf{D} \nabla \hat g \right ) - s \theta \hat g + m_0 + \hat r
\end{equation}
where $\hat g = \hat g(\emph{\textbf{x}},t,s) = L\{g(\emph{\textbf{x}},t,\tau), \tau, s\} = \int_0^{\infty} {e^{-s \tau} g(\emph{\textbf{x}},t,\tau) d\tau}$ is the transformed state of the function $g$, with $s \in \mathbb{Z}$ denoting the complex Laplace variable and $L\{\}$ the forward Laplace transformation operator, and where the transformed reaction term is $\hat r = q_\textrm{I}(\emph{\textbf{x}},t) - q_\textrm{O}(\emph{\textbf{x}},t) \hat g$. The zero-order source term $m_0 = m_0(\emph{\textbf{x}}, t) = \theta(\emph{\textbf{x}}, t) g(\emph{\textbf{x}}, t, \tau = 0)$ originates from the definition of the Laplace transform of the derivative of a function, i.e. $L\{\frac{\partial g(\emph{\textbf{x}},t,\tau)}{\partial \tau}, \tau, s\} = s \hat g(\emph{\textbf{x}},t,s) - g(\emph{\textbf{x}}, t, 0)$. Equation~(\ref{eq:TMLT_Eqn}) has the form of a standard transient-state advection-dispersion equation with first-order decay term (term $- s \theta \hat g$). The complex function $\hat g(\emph{\textbf{x}},t,s)$ is thus transported in space and time.\protect\\

The full set of transformed initial and boundary conditions yielding solution of equation~(\ref{eq:TMLT_Eqn}) can be formalized by:
\begin{subequations}\label{eq:TMLT_BCs}
\begin{equation}\label{eq:IC}
\hat g(\emph{\textbf{x}},0,s) = \hat f_0(\emph{\textbf{x}},s) \quad\quad \text{ in } \quad \Omega
\end{equation}
\begin{equation}\label{eq:type1}
\hat g(\emph{\textbf{x}},t,s) = 1 \quad\quad \text{on} \quad \Gamma_1
\end{equation}
\begin{equation}\label{eq:type2}
- \theta \textbf{D} \nabla \hat g \cdot \textit{\textbf{n}} = \hat{f}_2(t,s) \quad\quad \text{on} \quad \Gamma_2
\end{equation}
\begin{equation}\label{eq:type31}
\hat{\textbf{J}} \cdot \textit{\textbf{n}} = 0 \quad\quad \text{on} \quad \Gamma_0
\end{equation}
\begin{equation}\label{eq:type32}
\hat{\textbf{J}} \cdot \textit{\textbf{n}} = \textbf{q} \cdot \textit{\textbf{n}} \quad\quad \text{on} \quad \Gamma_3
\end{equation}
\end{subequations}
The function $\hat{f}_0(\emph{\textbf{x}},s)$ is a given function representing the age distribution at initial time $t=0$, the function $\hat{f}_2$ is used to represent a purely diffusive production of age mass along $\Gamma_2$ boundary portions, and the total age mass flux is $\hat{\textbf{J}} = \textbf{q} \hat g - \theta \textbf{D} \nabla \hat g$. In the Laplace domain the delta distribution $\delta(\tau)$ being simply 1, for all times the zero-age boundary condition is proportional to the net fluid flux $\textbf{q}(\emph{\textbf{x}},t) \cdot \textit{\textbf{n}}$. All types of boundary conditions can be handled in a standard way, just as is done by solute transport codes (given the particularities of each of the finite element, finite volume or finite differences methods in relation to the implementation of boundary conditions). The only difference is that boundary functions are expressed with complex numbers. However, they still can be treated using the same formalism than the one used with real arithmetics functions, because in fact the imaginary part of the zero-age condition is always zero. The Dirichlet condition is represented by the complex function $z = 1 + j \times 0$ (where $j$ denotes the imaginary unit of complex numbers, $j^2 = -1$), and the Cauchy condition is represented by the complex function $z = \textbf{q} \cdot \textit{\textbf{n}} + j \times 0$. Both conditions have a zero imaginary part and thus fall into the space of real numbers.\protect\\

The zero-order source term $m_0$ corresponds to an initial condition for $g$ on $\tau$ for all times $t$ and positions $\emph{\textbf{x}}$, and will need specific handling since it is not known and not necessarily zero at any time $t$, a priori. The only location where $m_0$ is known is along the inflow boundary $\Gamma_1$, where it is infinity. This implies that equation~(\ref{eq:TMLT_Eqn}) is of nonlinear type because it partially depends on the unknown solution $g(\emph{\textbf{x}}, t, 0)$. The term $m_0$ should thus not be neglected, in particular at the neighborhood of the inflow boundaries, and above all for hydrogeological configurations that tend to evolve toward increased mixing of ages or that contain important fractions of young age masses due to important infiltrations by superficial recharge. Neglecting this term would exclude the correct treatment of the distributions $g(\emph{\textbf{x}}, t, \tau)$ for which the maximum intensity is located around small values of $\tau$, as can be found with cases of young and shallow flow systems undergoing recharge by rainfall infiltration, or also with strongly dispersive systems inducing high mixing of age masses. The specific, however still very theoretical case of systems described by the exponential age distribution model corresponds to the worst configuration for which neglecting the term $g(\emph{\textbf{x}}, t, 0)$ would not allow the solution $g$ to show maximum age intensities at exactly $\tau = 0$. The numerical treatment of this source term will be discussed later in section~\ref{sec:TMLTG_CompuAspects}.\protect\\

\begin{figure}[h]
\noindent \includegraphics[width=0.5\textwidth]{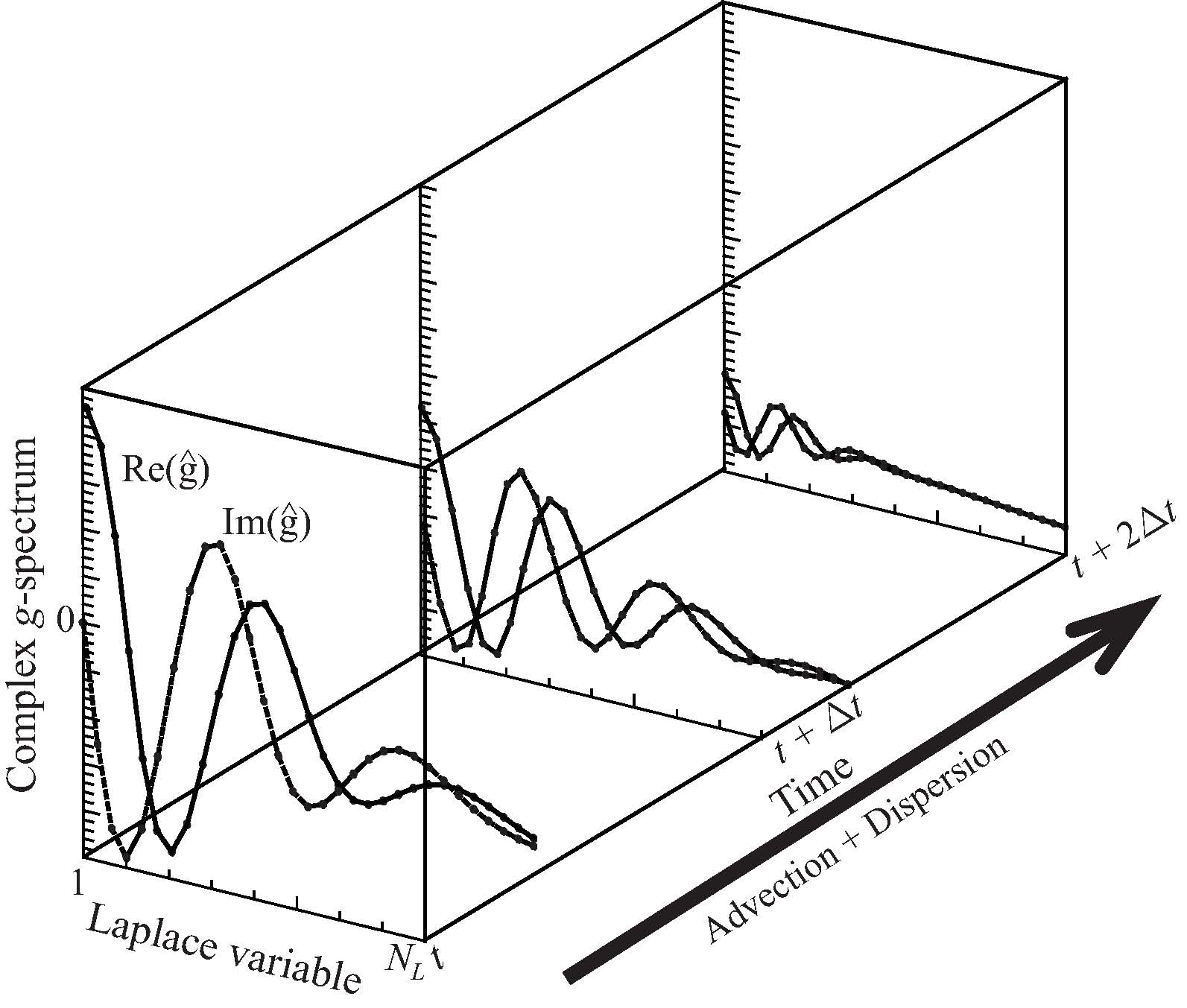}
\caption{Schematic representation of the TMLT technique used to solve the 5-D age equation.\label{fig:fig2}}
\end{figure}

Provided an initial distribution $g(\emph{\textbf{x}}, 0, \tau)$ and the prescription of appropriate boundary conditions, equation~(\ref{eq:TMLT_Eqn}) can be solved for a finite number of discrete values $s_k$ of the Laplace variable $s$, just as if the system of equations invoking equation~(\ref{eq:TMLT_Eqn}) for each discrete value $s_k$ would correspond to a multispecies mass transport equation system, with each of the species being assimilated to a specific discrete value $s_k$, and within which the species would not interact. By letting $\mathcal{L}_k() = \nabla \cdot \textbf{q} () - \nabla \cdot \theta \textbf{D} \nabla () + s_k \theta () + q_\textrm{O}()$ be the linear transport operator associated to variable $s_k$, the system of equations corresponding to $N_L$ discrete Laplace variables results in:
\begin{equation}\label{eq:TMLTSystem}
\left ( \frac{\partial \theta}{\partial t} + \mathcal{L}_k \right ) \hat g(\emph{\textbf{x}},t,s_k) =  \hat m_k \quad,\quad\quad k = 1,\ldots,N_L
\end{equation}
with $\hat m_k = q_\textrm{I} \hat g_I(s_k) + m_0$. The function $\hat g_I(s_k)$ is an age distribution expressed in the Laplace domain. It is Dirac for the classical case ($\hat g_I = 1$) or it needs to be specified as $\hat g_I(\emph{\textbf{x}}, t, s_k)$ for cases where internal volumes of water are produced with particular, not necessarily nil age distributions. Equation~(\ref{eq:TMLTSystem}) combines a time-evolution of Laplace transformed functions, and will consequently be referred to as the "Time-Marching Laplace-Transform" approach (TMLT). The transport process of complex functions operated with the TMLT approach is schematized in Figure~\ref{fig:fig2}. For fixed $\emph{\textbf{x}}$ and $t$, the function $\hat g(\tau)$ typically is an oscillatory function whose period increases with $s_k$ and depends on the model coefficients $\textbf{q}$ and $\textbf{D}$ (\cite{Sudicky89}).

The TMLT approach can also be used to solve for constituent exposure time density only for simplified cases where the advection velocity in the $\tau-$direction $\upsilon_{\alpha}$ and the reaction term $r_{\alpha}$ require a total independence on the independent variable $\tau$ and on the dependent variable $\rho_{\alpha}(\emph{\textbf{x}}, t, \tau)$. Another acceptable situation is when the product $\upsilon_{\alpha}(\emph{\textbf{x}}, \tau) \rho_{\alpha}(\emph{\textbf{x}}, t, \tau)$ and the reaction term $r_{\alpha}$ can both be transformed in the Laplace space. Given these restrictions, the governing equation to be solved is:
\begin{equation}\label{eq:5Ddensity_upst}
\frac{\partial \hat \rho_{\alpha}}{\partial t} + \nabla \cdot \left ( \textbf{v} \hat \rho_{\alpha} - \textbf{D} \nabla \hat \rho_{\alpha} \right ) + L\{\frac{\partial \upsilon_{\alpha} \rho_{\alpha}}{\partial \tau}, \tau, s\} = \hat r_{\alpha}
\end{equation}
Possible applications of equation~(\ref{eq:5Ddensity_upst}) can be met in problems defined by an exposure-time velocity $\upsilon_{\alpha}$ that follows a finite memory model (for example see the bacterial transport case presented by~\cite{Ginn00b}), with which $\upsilon_{\alpha}$ is represented by a left-continuous Heaviside step function ($H(y) = 0$ for $y \leq 0$ and $H(y) = 1$ for $y > 0$), $\upsilon_{\alpha}(\tau) = - H(\tau)$, with the consequence that the transformation of the term $\frac{\partial H(\tau) \rho_{\alpha}(\tau)}{\partial \tau}$ gives $s \hat \rho_{\alpha}$. \protect\\

To the author's knowledge, there is no such approach that lets time-marching processes operating on complex Laplace transformed functions, allowing solution of 5-dimensional equations of the form of equation~(\ref{eq:5Ddensity}).

\subsection{Numerical Inversion of the Transformed Functions}\label{sec:lapin}
To return into the real age-domain, the fields $\hat g(\emph{\textbf{x}},t,s)$ or $\hat \rho_{\alpha}(\emph{\textbf{x}}, t, s)$ need to be numerically inverted. Various authors (e.g. see \cite{Sudicky89,Cornaton03,Cornaton06a,Morales09}) have shown that the numerical inversion of Laplace Transform finite element solutions are generally accurate when using the algorithm of Crump (\cite{Crump76}) and when further combining it to the quotient-difference algorithm, as proposed by~\cite{deHoog82}. The Laplace Transform schemes can suffer from oscillations near discontinuities during the inversion, in relation to the non-uniform convergence of the series in the inversion formula. A control of these oscillations can be done by increasing the number of discrete Laplace variables $N_L = 2 n + 1$ and decreasing the mesh/grid size near discontinuities. It can be reported that a value for $n$ ranging between 10 and 20 generally yields accurate solutions in the presence of sharp gradients (\cite{Cornaton03,Cornaton06a,Morales09}).\protect\\

It can also be noted that the TMLT method provides not only a solution for the density distribution, but also for its integral (i.e. the cumulative distribution $G(\emph{\textbf{x}}, t, \tau) = \int_0^{\tau} {g(\emph{\textbf{x}}, t, u) d u}$), when one is taking advantage of the Laplace Transform property $\hat G(\emph{\textbf{x}}, t, s) = g(\emph{\textbf{x}}, t, s) / s$. Consequently, at any selected time $t$ it is possible to obtain both $g(\emph{\textbf{x}}, s)$ and $G(\emph{\textbf{x}}, s)$ by applying the inverse transform on $\hat g$ and $\hat G$.

\subsection{Computational Aspects of the TMLT Methods}\label{sec:TMLTG_CompuAspects}
In terms of memory usage, the TMLT method involves storage requirements that are proportional to $N_{\emph{\textbf{x}}} N_t N_L \ll N_{\emph{\textbf{x}}} N_t N_{\tau}$ ($N_x$, $N_t$, $N_{\tau}$ and $N_L$ being the number of computational points, computational times, age classes and Laplace variables, respectively). In most of the cases, the amount of data needed to store the age distributions in the complex Laplace space will be much smaller than the amount needed in the real age domain, rendering the method particularly interesting when very old ages are present in the modelled system. This amount of discrete data partly depends on the shape and complexity of the functions (e.g. multi-modal functions would obviously require a larger amount of discrete time values to allow for a good representation of the functions), but also on the maximum age that is to be accounted for. The bigger the maximum age is, the bigger the amount of age classes needed to properly represent the age distribution. This limitation is not affecting the TMLT method since for any extent of the age dimension the number of needed discrete Laplace variables remains unchanged. The method is consequently particularly well-suited for real-site applications that require fine discretization of the computational mesh/grid. \protect\\
The global computational demand of the TMLT method can be compared to that of a $N_L$-species mass transport problem, with the important simplification that the $N_L$ species are independent and do not react. The solving of the $N_L$ assembled algebraic systems requires iterative solvers that are capable of working with complex-arguments matrices and double-precision arithmetics. Even if the memory storage in the $\tau-$axis is highly reduced when working in the Laplace domain, the overall computational demand calls for parallelism and distributed/shared-memory calculations for handling real-site 3-D applications over large time periods.\protect\\

The zero-order mass source term $m_0$ in equation~(\ref{eq:TMLT_Eqn}) requires the knowledge of the spatial distribution of the density associated to age $\tau = 0$, since this density will not be zero for many system configurations, as previously discussed in section~\ref{sec:TMLTScheme}. The consequence is that the system of $N_L$ equations as formalized with equation~(\ref{eq:TMLTSystem}) is of nonlinear type. This induces an additional computational burden since such a nonlinearity has to be treated using iterative techniques such as the Picard or Newton-Raphson methods to carry out nonlinear iterations within each time-step $\Delta t$ to yield solution at time $t + \Delta t$. With these classical iterative methods the spatial distribution $g_0 = g(\tau = 0)$ at previous time $t$ is used as initial guess to proceed with a series of nonlinear iterations with which the field of $g_0$, and thus of $m_0$, is updated at each new nonlinear iteration in order to obtain the solution at time $t + \Delta t$. The spatial distribution of $g_0$ thus needs to be calculated by applying a numerical inverse transform to the function $\hat g(s)$ at the age value $\tau = 0$ after each nonlinear iteration, and until an appropriate stopping criterion indicates that convergence has been achieved. Convergence criteria will typically correspond to a threshold limit applied to the residues between two nonlinear iterations, and can be formulated e.g. using the standard absolute error measure $\epsilon_i = \, \mid {g_0^{(i,j)} - g_0^{(i,j-1)}} \mid$, where $j$ denotes the iteration level and $i$ the computational points index. However, the corresponding computational burden remains relatively minor for a priori only the immediate vicinity of the inflow boundary is concerned. \protect\\
As an alternative to the iterative solution, a simplistic approach is to assume an \emph{explicit scheme} for calculating the state of the function $g_0$ at time $t + \Delta t$ from its previous state at time $t$, i.e. by letting $g(\emph{\textbf{x}}, t + \Delta t, \tau = 0) \simeq g(\emph{\textbf{x}}, t, \tau = 0)$. Of course the explicit approach will hardly converge nor produce stable results with stiff problems, unless it is associated to small time-steps $\Delta t$ to keep the error in the result bounded, which in turn may render the solution impractical.\protect\\

Nonlinearities such as the ones induced by the dependence of the flow equation to saturation or fluid density are implicitly handled by the TMLT method. For each time-level, variable saturation conditions can be updated, as well as fluid density $\rho$ and viscosity $\mu$, yielding time- and space-dependent moisture content $\theta = \theta(\emph{\textbf{x}}, t)$ and velocity $\textbf{q} = \textbf{q}(\theta, \rho, \mu, t)$. The age equation problem is then solved with the corresponding distribution of $\theta$ and $\textbf{q}$. Concerning the viscosity and density dependency of fluid flow, the coupled flow and nonlinear thermohaline problems can e.g. be solved at a given time-level, preliminary to solving for the age problem and go forward to the next time-level. Even if in the present work only velocity fields derived from the incompressible groundwater flow problem are considered, it can be noted that the method is also suited for operating with compressible homogeneous fluid flow problems. In addition, we can also argue that the TMLT method does not present any restrictive limitation in relation to its adaptation to coupled surface-subsurface flow problems, since an integrated and implicit flow solution can simply be performed at any time-level and the resulting velocity field be used to solve for the age equation at the same time-level. Unstable TMLT solutions (e.g. originating from advection-dominated transport processes) can also be supported without any specific restriction by stabilization methods (such as Petrov-Galerkin based formulations), or upstream-weighting and mass-lumping formulations.\protect\\

When discretization techniques are used to solve with the TMLT method, the terminology can be refined. For instance, when the finite element technique is applied, the discretization of the system of equations~(\ref{eq:TMLTSystem}) yields the TMLTG technique ("Time-Marching Laplace-Transform Galerkin"), in reference to the works of~\cite{Sudicky89} on the LTG technique. When the finite-volume technique is applied, one may refer to the TMLTFV technique. The finite differences technique would equivalently yield the TMLTFD technique.

\subsection{The TMLTG Solution Technique}\label{sec:TMLTG}
In this section, we apply the finite element framework in order to illustrate a numerical implementation of the TMLT technique, and to provide additional details related to the computational burden associated to the discretization of equation~(\ref{eq:TMLT_Eqn}).

\subsubsection{Finite Element Formulation}\label{sec:TMLTG_FE}
We consider the specific case of the transient groundwater age distribution equation for one value of the Laplace variable $s_k$:
\begin{equation}\label{eq:TMLTEqSk}
\frac{\partial \theta \hat g_k}{\partial t} + \nabla \cdot \left ( \textbf{q} \hat g_k - \theta \textbf{D} \nabla \hat g_k \right ) + \hat \Theta_k \hat g_k = m
\end{equation}
with $\hat \Theta_k = q_\textrm{O} + s_k \theta$ and $m = q_\textrm{I} + \theta g(\emph{\textbf{x}}, t, 0)$. Initial conditions
and standard boundary conditions yielding solutions of
equation~(\ref{eq:TMLTEqSk}) are given by equation~(\ref{eq:TMLT_BCs}).\protect\\

When a standard Galerkin finite element formulation is applied to equation~(\ref{eq:TMLTEqSk}), the following discretized equation is obtained:
\begin{align}\label{eq:TMLTGalerkinEq}
 & \int_\Omega {{\textbf{N}}(\frac{{\partial \theta \hat \psi_k }}{{\partial t}} + \hat \Theta_k \hat \psi_k)d\Omega} \\
 - & \int_\Omega {\nabla {\textbf{N}}({\textbf{q}}\hat \psi_k - \theta {\textbf{D}}\nabla \hat \psi_k )d\Omega} \nonumber \\
 - & \int_{\Gamma_+} {{\textbf{N}}(\theta {\textbf{D}}\nabla \hat \psi_k - {\textbf{q}}\hat \psi_k) \cdot \emph{\textbf{n}} d\Gamma} \nonumber \\
 = {} & \int_\Omega {{\textbf{N}} m d\Omega} + \int_{\Gamma_3} {{\textbf{N}}({\textbf{q}}\hat \psi_k - \theta {\textbf{D}}\nabla \hat \psi_k) \cdot \emph{\textbf{n}} d\Gamma} \nonumber
\end{align}
where the vector $\textbf{N} = [N_1,...,N_{nn}]$ is the element shape functions vector, that is equal to the weighting functions in a standard Galerkin formulation, and where the unknown variable is denoted by $\hat \psi_k = \hat \psi_k(\emph{\textbf{x}}, t, s_k)$. Note that the third term in the L.H.S. of
equation~(\ref{eq:TMLTGalerkinEq}) exhibits the gradient projection of the
dependent variable along open boundaries $\Gamma_+$. If not neglected, this term will require specific handling, e.g. using the technique proposed by~\cite{Frind88} or~\cite{Cornaton04}.\protect\\
On any finite element domain $\Omega_e$, the unknown variable is replaced by a continuous approximation of the form $\hat \Psi_k = \sum_{n=1}^{nn} {\hat \psi_{k,n} N_n}$, where $\hat \psi_{k,n}$ are the $nn$ nodal unknowns. After substitution of
this trial solution into~(\ref{eq:TMLTGalerkinEq}), and after dividing the domain and
boundary integrals into piecewise elemental contributions over an
element $\Omega_e$, the following global matrix system of algebraic equations is obtained by summing the $ne$ elemental
matrices:
\begin{equation}\label{eq:TMLTGalerkinFEeq}
[\hat{\textbf{A}}_k] \{ \hat \Psi_k \} = \{ \hat {\textbf{F}}_k \} - [\textbf{M}] \{ {\frac{\partial
\hat \Psi_k}{\partial t}} \}
\end{equation}
where $[\hat{\textbf{A}}_k]$ is the global (complex) stiffness matrix, $[\textbf{M}]$ the
mass matrix, and $\{ \hat {\textbf{F}}_k \}$ the load vector:
\begin{subequations}\label{eq:TMLTGalerkinFEMat}
\begin{equation}\label{eq:TMLTGalerkinFEMat1}
[\hat{\textbf{A}}_k] = \sum_{e=1}^{ne} {\int_{\Omega_e} {( \textbf{B}
\theta \textbf{D} \textbf{B}^T - \textbf{B} \textbf{q}} \textbf{N}^T +
\textbf{N} \hat \Theta_k \textbf{N}^T) d\Omega_e}
\end{equation}
\begin{equation}\label{eq:TMLTGalerkinFEMat2}
[\textbf{M}] = \sum_{e=1}^{ne} {\int_{\Omega_e} {\textbf{N} \theta
\textbf{N}^T d\Omega_e}}
\end{equation}
\begin{equation}\label{eq:TMLTGalerkinFEMat3}
\{ \hat {\textbf{F}}_k \} = \sum_{e=1}^{ne} {\int_{\Gamma_e} {\textbf{N} (
\textbf{q} \textbf{N}^T - \theta \textbf{D} \textbf{B}^T ) \cdot
\emph{\textbf{n}}d\Gamma_e} + \int_{\Omega_e} { \textbf{N}
m \textbf{N}^T d\Omega_e}}
\end{equation}
\end{subequations}
with the gradient matrix $\textbf{B} = \nabla \textbf{N}$.\protect\\

The linear algebraic system~(\ref{eq:TMLTGalerkinFEeq}) needs to be assembled and solved $N_L$ times, i.e. for each independent Laplace variable $s_k$. However, for one given time-level the mass matrix $[\textbf{M}]$ is real and only needs to be assembled once. The complex stiffness matrix $[\hat{\textbf{A}}_k]$ can be splitted into a real part $[\textbf{A}^r]$ and a complex part $[\hat{\textbf{A}}^c_k]$, with:
\begin{equation}
[\textbf{A}^r] = \sum_{e=1}^{ne} {\int_{\Omega_e} {(\textbf{B} \theta \textbf{D} \textbf{B}^T - \textbf{B} \textbf{q} \textbf{N}^T + \textbf{N} q_\textrm{O} \textbf{N}^T) d\Omega_e}} \nonumber
\end{equation}
and
\begin{equation}
[\hat{\textbf{A}}^c_k] = \sum_{e=1}^{ne} {\int_{\Omega_e} {\textbf{N} s_k \theta \textbf{N}^T d\Omega_e}} \nonumber
\end{equation}
The real matrix only requires a single evaluation, while the complex matrix needs evaluation for each Laplace variable $s_k$. However, the complex elementary matrix $[\hat{\textbf{A}}^c_k]_e$ is ideally obtained by scaling the real elementary mass matrix $[\textbf{M}]_e = \int_{\Omega_e} {\textbf{N} \theta \textbf{N}^T d\Omega_e}$ by $s_k$, $[\hat{\textbf{A}}^c_k]_e = s_k [\textbf{M}]_e$. Finally, since the zero-age boundary condition is always 1 in the Laplace space, the load term $\{ \textbf{F}_k \}$ remains identical for all the $N_L$ Laplace variables, and also needs to be assembled only once for a given time-level, $\{ \hat {\textbf{F}}_k \} = \{ \textbf{F} \}$. Consequently, the computational burden for the assembling of the $N_L$ equations is relatively low, and once the real-arguments matrices and vectors $[\textbf{A}^r]$, $[\textbf{M}]$ and $\{ \textbf{F} \}$ have been evaluated for the first Laplace variable $s_1$, the remaining $N_L - 1$ assemblings are obtained by scaling $[\textbf{M}]$ by $s_k$ and by adding the contribution of the accumulation term $- [\textbf{M}] \{ {\frac{\partial
\hat \Psi_k}{\partial t}} \}$ to $\{ \textbf{F} \}$. Finally note that the matrix operations required for assembling and solving the linear algebraic system (16) are the same for complex matrices as they are defined for matrices with real entries, such that the complexity of (16) does not add any particular additional technical difficulty.

\subsubsection{Time Discretization}\label{sec:TMLTG_TimeMarching}
The well-known numerical artifacts induced by badly controlled time-marching schemes affecting standard advection-dispersion transport solutions also affect the TMLT method in a similar way. This implies that the same constraints regarding the time-discretization have to be considered to avoid numerical instabilities (like the constraints on the P\'{e}clet and Courant numbers), and that appropriate adaptive time-stepping solution techniques also need to be designed. The accuracy of the TMLT solutions will depend on the adopted time integration scheme and stepsize, exactly as with standard transient advection-dispersion equations. Whenever the choice on the stepsize is not appropriate to ensure stability and/or accuracy, the solutions will have the tendency of being affected by numerical diffusion when fully implicit schemes are used or by numerical oscillations when fully explicit or centered schemes are used.\protect\\
The time-marching procedure needed to solve the transient linear algebraic system~(\ref{eq:TMLTGalerkinFEeq}) can be obtained by a standard finite-differences approximation of the time-derivatives. The resulting
algebraic system can be formalized by the following:
\begin{equation}\label{eq:TMLTGTimeMarchingEq}
\{ \hat \Psi_k \}^{t+\Delta t} = [\hat{\textbf{C}}_k]^{-1} \{ \hat \Psi_k \}^{t} + \{
\hat{\textbf{Q}}_k \}^t
\end{equation}
with coefficient matrix
\begin{equation}\label{eq:CMatrix}
[\hat{\textbf{C}}_k] = \left ( \varepsilon [\hat{\textbf{A}}_k]^{t + \varepsilon
\Delta t} + \frac{1}{\Delta t}[\textbf{M}]^{t + \varepsilon \Delta
t} \right )
\end{equation}
and accumulation term
\begin{equation}\label{eq:QVector}
\{\hat{\textbf{Q}}_k \}^t = \left ( (\varepsilon - 1) [\hat{\textbf{A}}_k]^{t +
\varepsilon \Delta t} + \frac{1}{\Delta t}[\textbf{M}]^{t +
\varepsilon \Delta t} \right ) \{ \hat \Psi_k \}^t + \{ \textbf{F} \}^{t +
\varepsilon \Delta t}
\end{equation}
with $\varepsilon$ varying between 0 and 1. Like with standard transport scheme, when the explicit scheme is used ($\varepsilon = 0$) no iterations are needed since the matrices are evaluated only
with the initial value $\{\hat \Psi_k \}^t$, always constant over a given
time-step. However, to ensure stability, convergence
and accuracy, the Crank-Nicholson scheme ($\varepsilon =
0.5$) is generally recommended. The fully-implicit scheme
($\varepsilon = 1$) is known to yield smoother solutions in
relation to its intrinsic diffusive nature. Adaptive time-stepping strategies can be designed to automatically control the stepsize $\Delta t$. One can classically use the evolution of the norm of the complex spectrum $\hat g$ between two or more time-steps to predict an optimal stepsize.
%
\section{Validation and Illustration Examples}\label{sec:validation}
In this section the TMLTG scheme is validated and illustrated by means of analytical and numerical solutions. The numerical solutions have been performed with the \textbf{GW} software (\cite{Cornaton07b}), which contains the characteristics of the TMLTG method as described in section~\ref{sec:TMLTG}. The nonlinear iterations induced by the zero-order source term $m_0(\emph{\textbf{x}}, t)$ in equation~(\ref{eq:TMLT_Eqn}) are performed using the Newton-Raphson iterative scheme. Initial conditions for the distribution $g$ are obtained by solving the steady-state equation~(\ref{eq:4Dage}) using the LTG formalism~(\cite{Sudicky89,Cornaton03}).

\subsection{One-Dimensional Validation of the TMLTG Numerical Solution}\label{sec:1dtransient}
\begin{figure*}[h]
\noindent \includegraphics[width=0.9\textwidth]{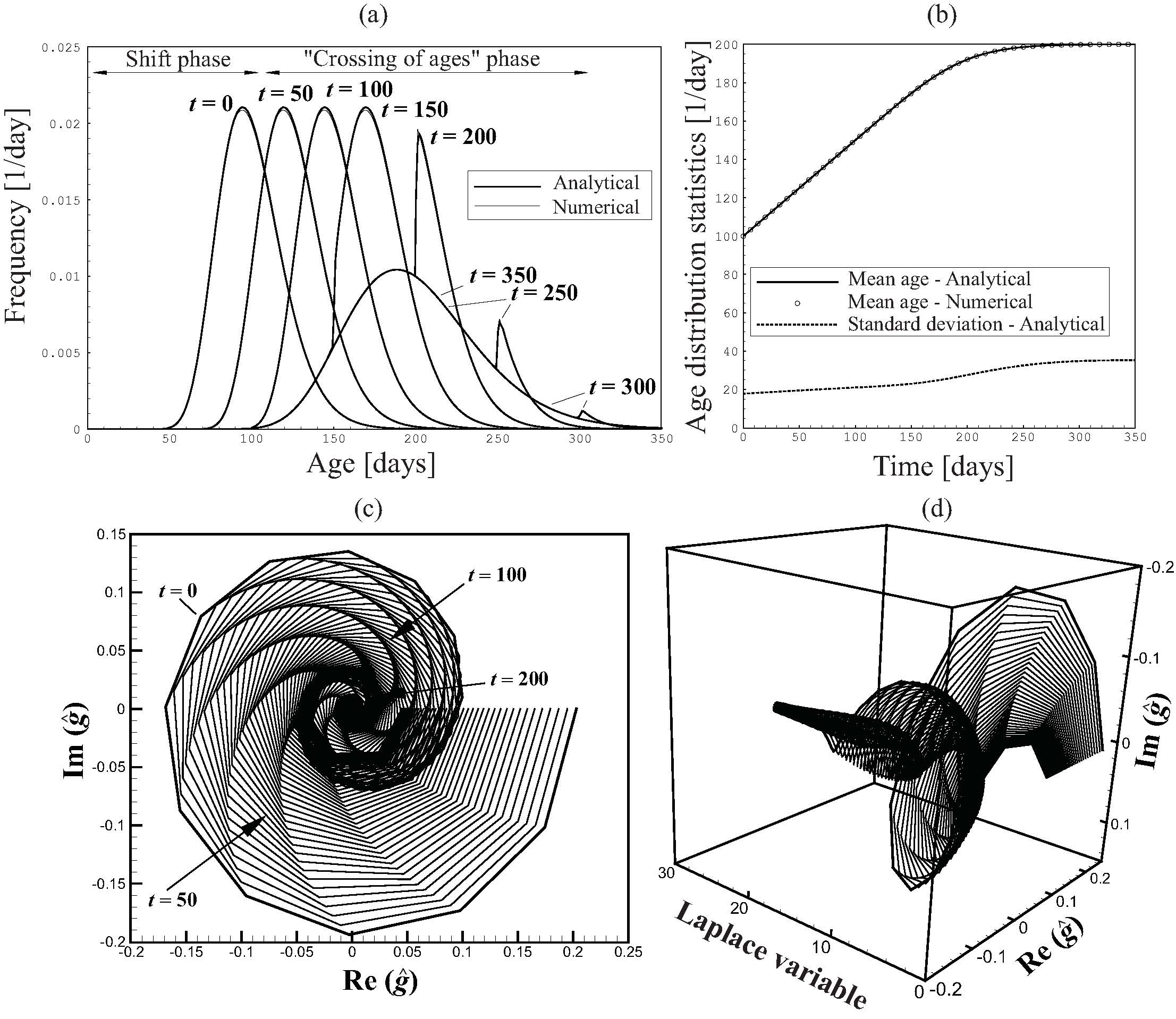}
\caption{Temporal evolution of the 1-D transient-state age
distribution $g(x,t,\tau)$ at $x = 100$~m. (a) Distribution $g(x,t,\tau)$ at some selected times; (b) Temporal moments: mean age $a(x,t)$ and standard deviation $\sigma_g(x, t)$; (c) and (d) Real and imaginary parts of $\hat g(s)$ at some selected times.\label{fig:fig3}}
\end{figure*}

The robustness of the TMLTG scheme is validated using the pseudo-analytical 1-D solution derived in the Appendix. This analytical solution is taken as reference solution by considering a finite domain of length $L = 200$ meters, discretized into 200 pipe elements of 1 meter length. Dispersivity and molecular diffusion are fixed to $\alpha_L = 2$~m and $D_m = 0$, respectively. An initial equilibrated age distribution $g(x,0,\tau)$ is considered as being the steady-state solution resulting from the uniform and constant velocity $v = 1$~m/d. At $t \geq 0$, this velocity is assumed to be two times lower, i.e. $v = 0.5$~m/d. The resulting temporal evolution of the age
distribution $g(x,t,\tau)$ is shown in Figure~\ref{fig:fig3}a at the position $x = 100$ meters. The figure compares the analytical and numerical solutions. The TMLTG numerical solution makes use of the Crank-Nicholson scheme for the time-discretization, with a constant stepsize of 0.05 days. The number of Laplace variables used to store the age distributions is $N_L = 2 n + 1 = 31$. Figure~\ref{fig:fig3}b shows the temporal evolution of mean age $a(x, t)$ and standard deviation $\sigma_g(x, t)$ at $x = 100$~meters. These two quantities have been obtained by postprocessing the transient age distribution. Note that this is a means for validating the TMLTG solution obtained for this example, since the direct mean age solution can be taken as a robust reference from a numerical point of view. Mean age is two times bigger for the new equilibrium (at initial state it equals $\frac{L}{v} = \frac{100}{1} = 100$~days, and at final state it equals $\frac{L}{v} = \frac{100}{0.5} = 200$~days), and so is standard deviation. With respect to the evolution in time of the age distribution, one can see that from initial time and during approximately one pore volume (i.e $t = \frac{x}{v} = \frac{100}{1} = 100$~days, the turnover time corresponding to the initial condition), the initial age distribution is simply shifted toward bigger age values, its shape remaining intact. After time $t = 100$~days, the distribution experiences modifications and displays a successive apparition of peaks, a reduction of its maximum intensity and the increase of its dispersion. This transition phase is occurring until the new equilibrium is met, after about 350 days (thus more than one initial pore volume plus one new pore volume), and it can be described as being the time-period during which a \emph{crossing of ages} occurs. This particular time-period corresponds to the phase during which the contribution of infiltrated water molecules under the new flow regime to the new state of the age distribution is occurring. It is clear that this contribution needs, at least, one pore volume to be effective. In other words, since velocity has been lowered by a factor 2, all water molecules need two times the amount of time to travel to a same downstream location in comparison with the initial (equilibrated) situation. Thus, during at least one (previous) pore volume the age distribution is shifted toward older ages. After, it mixes with older ages (the ones of molecules that took more than a pore volume to reach the same location) and it progressively transforms until new equilibrium is met.\protect\\
The surprising snail shell shape image displayed in Figure~\ref{fig:fig3}c corresponds to the time evolution of the complex age distribution (still in the Laplace space) at the position $x = 100$ meters. The 2-D graph on the figure plots the imaginary part of $\hat g(x,t,s)$ against its real part, from initial state ($t = 0$~days) to final state ($t = 350$~days), with a temporal resolution of 5 days. The 3-D graph in Figure~\ref{fig:fig3}d gives a similar space-time representation.

\begin{figure*}[h]
\noindent \includegraphics[width=0.9\textwidth]{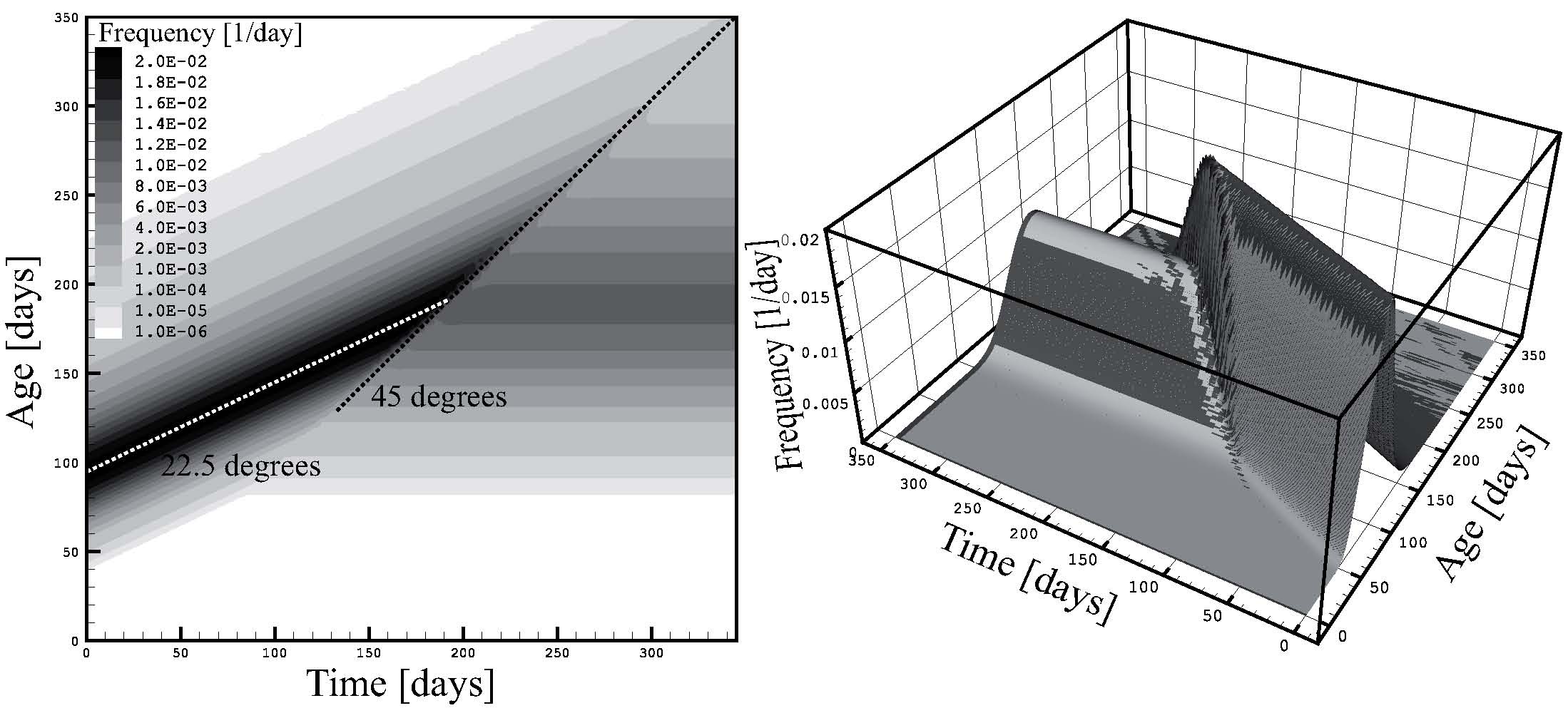}
\caption{Temporal evolution of the 1-D transient-state age
distribution $g(x,t,\tau)$ at $x = 100$~m. Left: 2-D representation; Right: 3-D representation showing the analytical solution in dark grey and the numerical solution in grey.\label{fig:fig4}}
\end{figure*}

Figure~\ref{fig:fig4} shows the same solutions using 2-D and 3-D representations. The shifting phase is well illustrated by an evolution of the age distribution in time following an angle of $\frac{90}{4} = 22.5$ degrees, showing that the increase of the maximum intensity of ages is of 0.5 per unit time, in relation to the new flow regime dictated by a two-times smaller velocity as compared to the previous flow regime (i.e. there is a ratio of 2 between previous and new velocity, or equivalently between the previous and new turnover time). This phase is followed by the progressive shaping of the new distribution (when ages from the previous state are crossed by ages of the new state), until the new equilibrium is met. The "crossing of ages" phase is characterized by an angle of exactly 45 degrees, reflecting the fact that all water molecules experience the same ageing per unit time.

\begin{figure*}[h]
\noindent \includegraphics[width=0.9\textwidth]{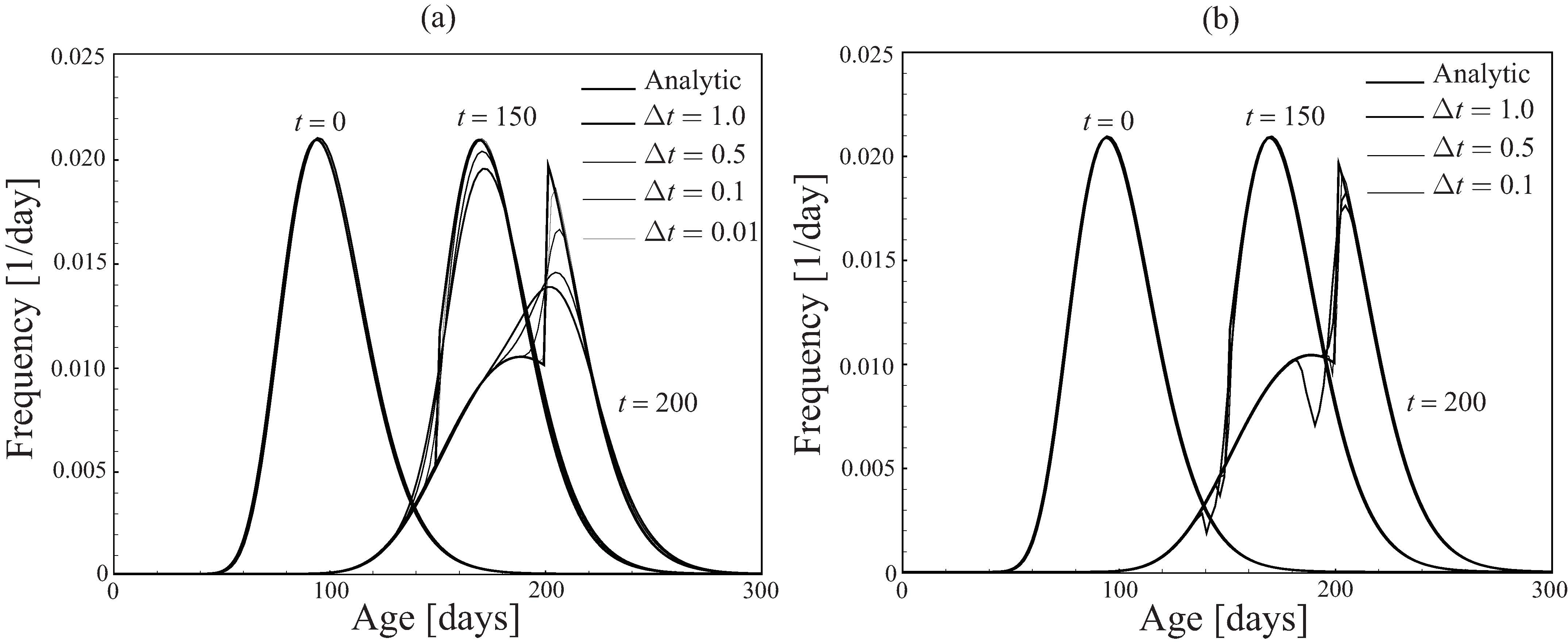}
\caption{Effects of the time-integration scheme and of the stepsize on the temporal evolution of the 1-D transient-state age
distribution $g(x,t,\tau)$ at $x = 100$~m. The figure compares the analytical solution to the numerical solutions obtained with the Crank-Nicholson and implicit time-integration schemes. (a) Implicit solution; (b) Crank-Nicholson solution. \label{fig:fig5}}
\end{figure*}

As discussed in section~\ref{sec:TMLTG_TimeMarching}, the accuracy and physical meaning of the TMLTG solutions are directly related to the time-marching scheme and the stepsize that are employed. We briefly illustrate the effect of the centered Crank-Nicholson and implicit schemes on the 1-D solution, by making use of some selected stepsizes. The results are shown in Figure~\ref{fig:fig5}, where the behavior of the solutions meets the classical, well-known effects of integration schemes on advective-dispersive transport solutions. With a stepsize of 1 day, the implicit scheme generates a very diffusive solution (in relation to the numerical diffusion that is artificially added by such a scheme) and the Crank-Nicholson scheme generates a highly oscillating solution. When the stepsize is progressively reduced, the oscillations with the Crank-Nicholson scheme are damped and the solution quickly converges to the exact solution (met for $\Delta t = 0.1$~days), while the implicit scheme becomes less diffusive and is getting closer to the exact solution (however not fully met for $\Delta t = 0.01$~days).

\subsection{Example of a 2-D Aquitard-Aquifer System with Natural Recharge}\label{sec:2Dexamples}
\begin{figure*}[h]
\noindent \includegraphics[width=0.9\textwidth]{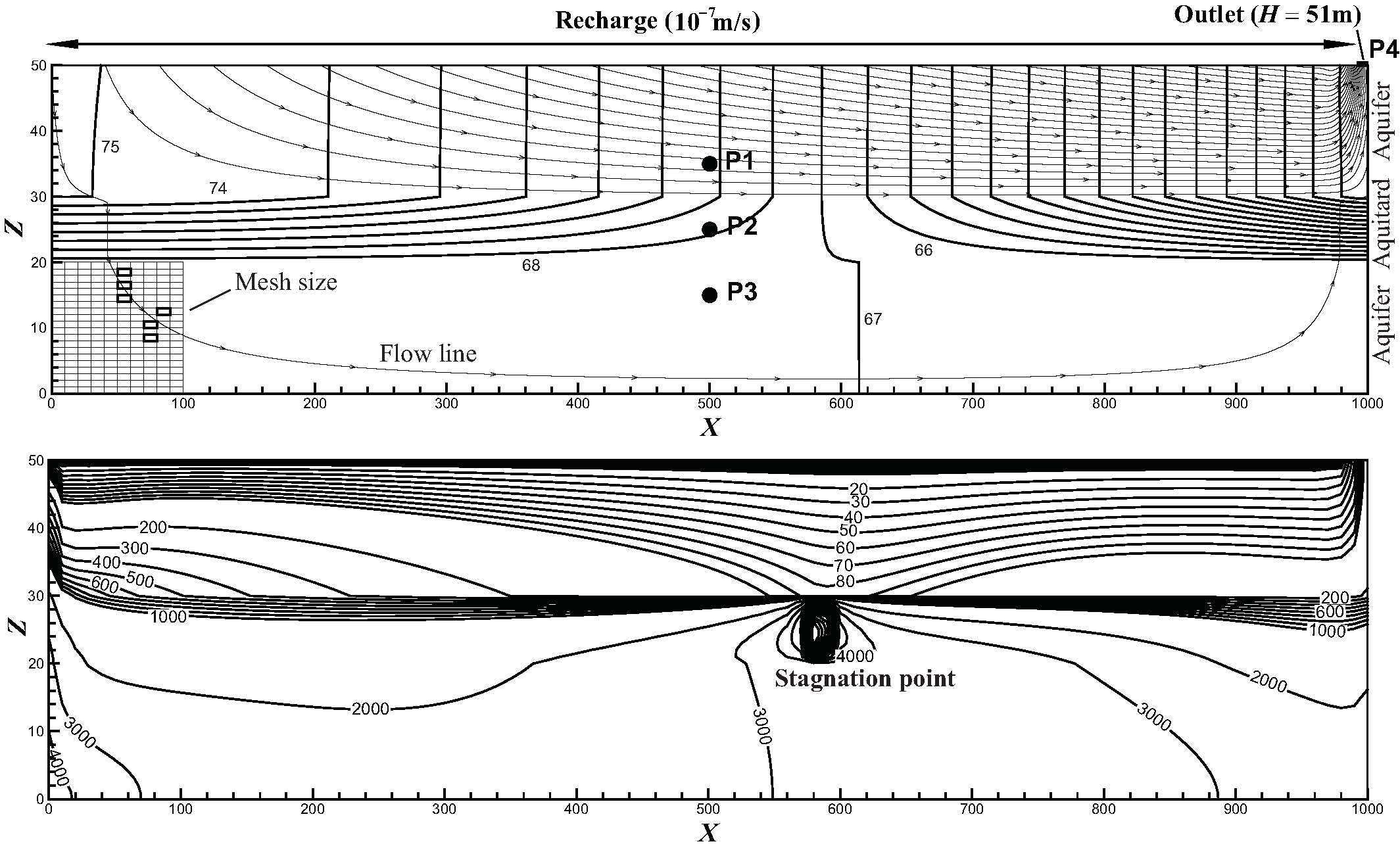}
\caption{Theoretical 2-D vertical aquitard-aquifer system with uniform recharge. (a) Hydraulic solution at steady-state with flow net illustrated by means of the advective tracking of some particles from recharge to discharge. (b) Mean age solution at steady state in days. The points P1, P2, P3 and P4 are observation points used to show the evolution in time of the hydraulic and age solutions. \label{fig:fig6}}
\end{figure*}

In order to illustrate the effect of annual recharge fluctuations on age distributions, we make use of a 2-D synthetic vertical system that corresponds to an idealization of a 1000 meters long and 50 meters thick aquitard-aquifer system, as shown in Figure~\ref{fig:fig6}. The 10-meters thick aquitard is located between the elevations $z = 20$~m and $z = 30$~m. The mesh is discretized using bilinear quadrangular elements of size $10 \times 1$ meters. Recharge is supposed to be uniformly distributed on top of the model, and is modelled by prescribing fluxes (Neumann-type condition). The outlet has a finite size of 10 meters, and is located at $z = 50$~m between the points $x = 990$~m and $x = 1000$~m. A fixed hydraulic head of 51 meters is prescribed. Hydraulic conductivity in the two aquifers is $K = 10^{-4}$~m/s, and aquitard hydraulic conductivity is $K = 10^{-8}$~m/s. The storage coefficient is $S = 7.5 \times 10^{-4}$~1/m in the aquifers and $S = 10^{-6}$~1/m in the aquitard. Aquifer porosity is 0.1, and aquitard porosity is 0.35. Dispersivity coefficients are nil in the aquitard, while longitudinal dispersivity is $\alpha_L = 10$~m and transverse dispersivity is $\alpha_T = 1$~m in the aquifers. Finally, both aquitard and aquifers have the same coefficient of molecular diffusion $D_m = 2.3 \times 10^{-9}~\textrm{m}/\textrm{s}^2$.\protect\\
The initial state is obtained by assuming a steady-state recharge of intensity $10^{-7}$~m/s (the average of the time-series applied at transient-state), and is represented in Figure~\ref{fig:fig6}. The hydraulic solution displays vertical drainage effects through the aquitard, downward oriented between approximately $x = 0$ and $x = 600$, and upward oriented after $x = 600$. The steady-state mean age distribution is typical of the one occurring in uniformly recharged systems (generating exponential-like age distributions), but it also displays the perturbation induced by the presence of the aquitard, which generates mixing of ages by diffusion.\protect\\

\begin{figure*}[h]
\noindent \includegraphics[width=0.9\textwidth]{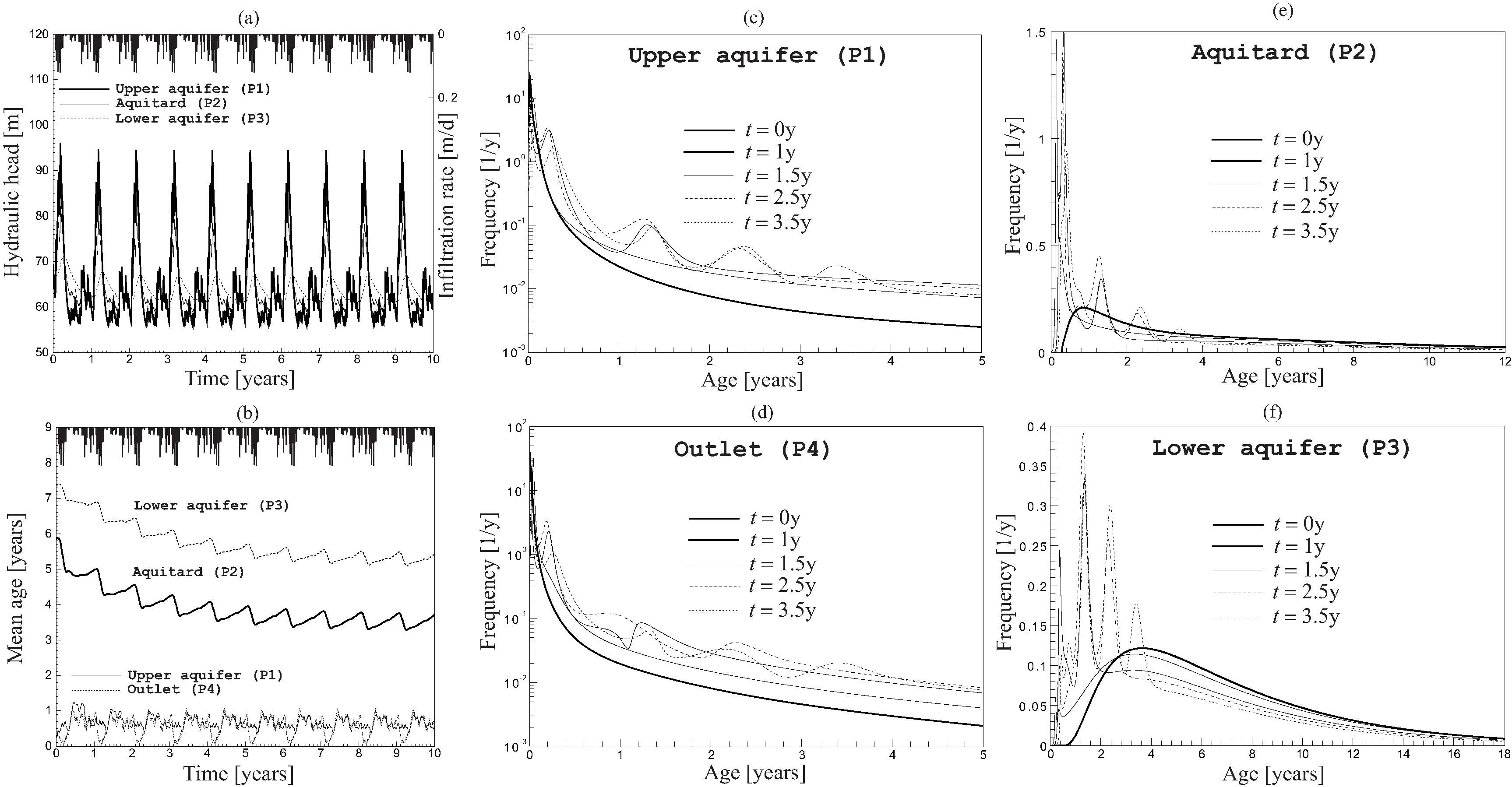}
\caption{Temporal evolution of flow, mean age and age distribution in the theoretical 2-D vertical aquitard-aquifer system of Figure~\ref{fig:fig6}. (a) Hydraulic head; (b) Mean age; (c) Age distribution above the aquitard; (d) Age distribution at the outlet; (e) Age distribution within the aquitard; (f) Age distribution under the aquitard. \label{fig:fig7}}
\end{figure*}

The transient-state scenario corresponds to the application of a cyclic forcing using a recharge rate of variable intensity on top of the model, and by keeping the water level at outlet unchanged. Simulations have been performed using the implicit time-integration scheme with a constant stepsize of 1 day, and using $N_L = 2 n + 1 = 41$ discrete Laplace variables. The transient evolution of hydraulic head and mean age at the observation points is given in Figures~\ref{fig:fig7}a and~\ref{fig:fig7}b. Mean age variations are high, in relation to the imposed high hydraulic variations. A first observation is related to the fact that the steady-state situation (using the average of annual fluctuations of infiltration) does not provide an initial condition that is close to some average, or pseudo-equilibrium, considering that transient fluctuations of infiltration are continuously repeated for several years. The new pseudo-equilibrium is met only after repeating annual series during at least 8 years for the aquitard and the lower aquifer, while the upper aquifer and the outlet tend to equilibrate around annual fluctuations already after 1 year. This observation points out the importance of evaluating realistic initial conditions for age mass transport problems, the steady-state (average) situation being not necessarily a proper representation of an initial state.
A general correlation with the hydrologic events can be done: the intuitive tendency of ageing of water during drought periods and vice-versa can be observed. This tendency however includes some delays between the hydrologic and age responses, especially in the isolated, confined lower aquifer and in the aquitard.\protect\\

\begin{figure*}[h]
\noindent \includegraphics[width=0.9\textwidth]{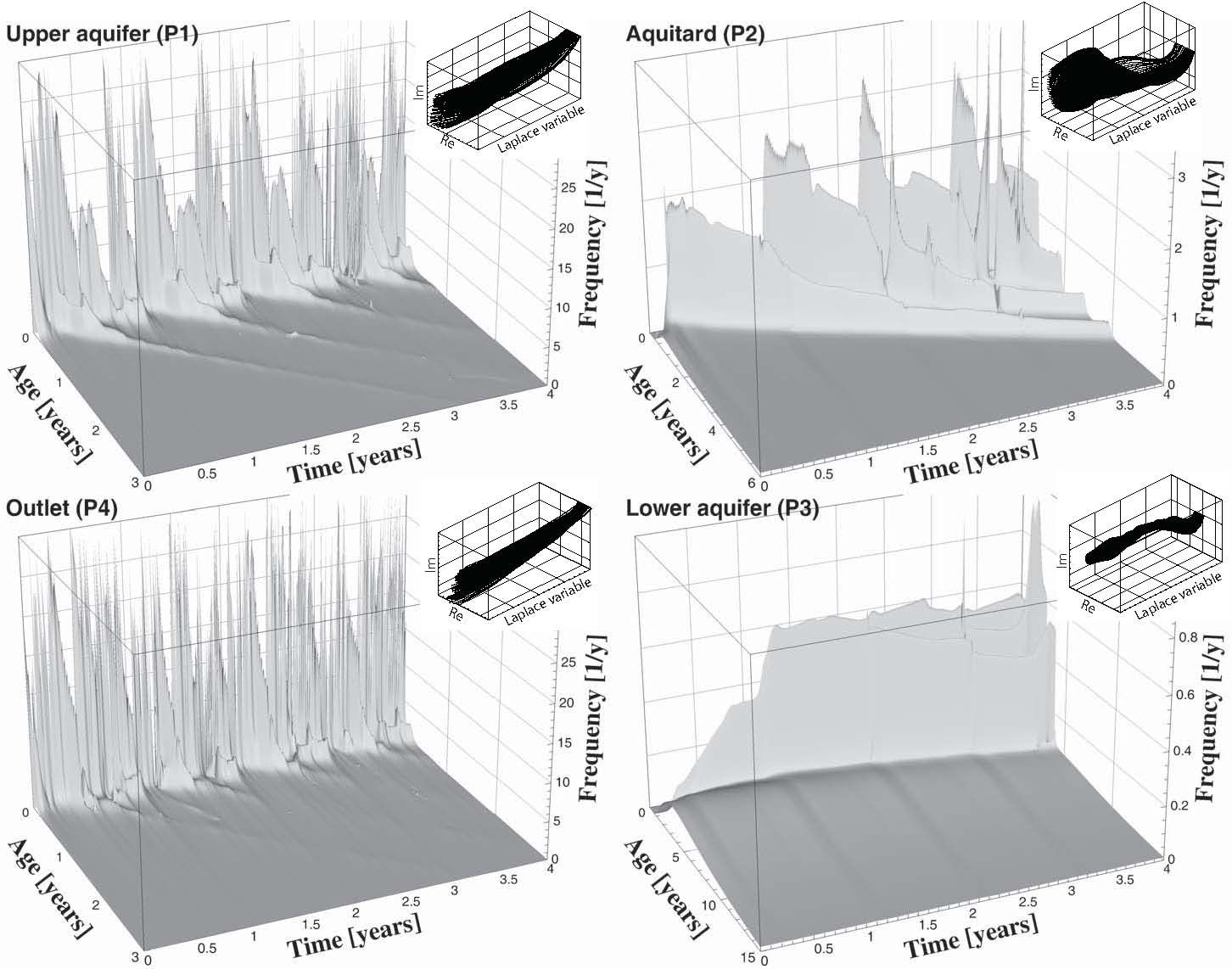}
\caption{Temporal evolution of the age frequency distributions in the theoretical 2-D vertical aquitard-aquifer system of Figure~\ref{fig:fig6}, using a surface representation.\label{fig:fig8}}
\end{figure*}

The transient age frequency distributions recorded at the observation points are given in Figures~\ref{fig:fig7}c to~\ref{fig:fig7}f and in Figure~\ref{fig:fig8}. The distributions display a relative complex temporal evolution as compared to the smooth aspect of the initial, steady-state distributions. All distributions become multi-modal with a successive generation of several modes (peaks). The aquitard and lower aquifer are invaded by young water due to new infiltration rates. This is revealed by the apparition of new modes at young ages. As time passes, these new water components take age and a displacement of these modes towards older age values can be observed, with an angle close to 45 degrees, as a reflection of the unit increase of age per unit time (i.e. the line passing by the maximum intensity of ages is defined by $\frac{\Delta \tau(g_{max})}{\Delta t} \simeq 1$). The annual new amount of freshly infiltrated water generates new peaks, and the consequence is that age distributions become multi-modal. The same observations can be made for the points P1 and P4 in the upper aquifer, with the difference that these points show much less tailing effect than points P2 and P3. The linear shape of the $\hat g$-spectrum for points P1 and P4 is typical of situations that are close to the conditions of the perfect exponential mixing (uniform recharge from the top along the entire domain). On the opposite, the $\hat g$-spectrum for points P2 and P3 differs from the linear, exponential-like case, and is more similar to that of the 1-D example presented in section~\ref{sec:1dtransient} (see Figure~\ref{fig:fig3}). An important observation is that, as opposed to the steady-state situation for points P1 and P4 (that both display the theoretical situation of an exponential mixing), the transient response does not present the same behavior but rather displays age distributions characterized by multi-modality and values at the origin that strictly differ from the exponential case (with which $g(\tau = 0) \simeq 1 / \tau_0$, with $\tau_0$ denoting the turnover time, or ratio between porous volume and steady-state discharge rate).

\subsection{Example of a Pumping-Well in a 3-D Homogeneous Aquifer with Natural Recharge}\label{sec:3Dexamples}
\begin{figure*}[h]
\noindent \includegraphics[width=0.9\textwidth]{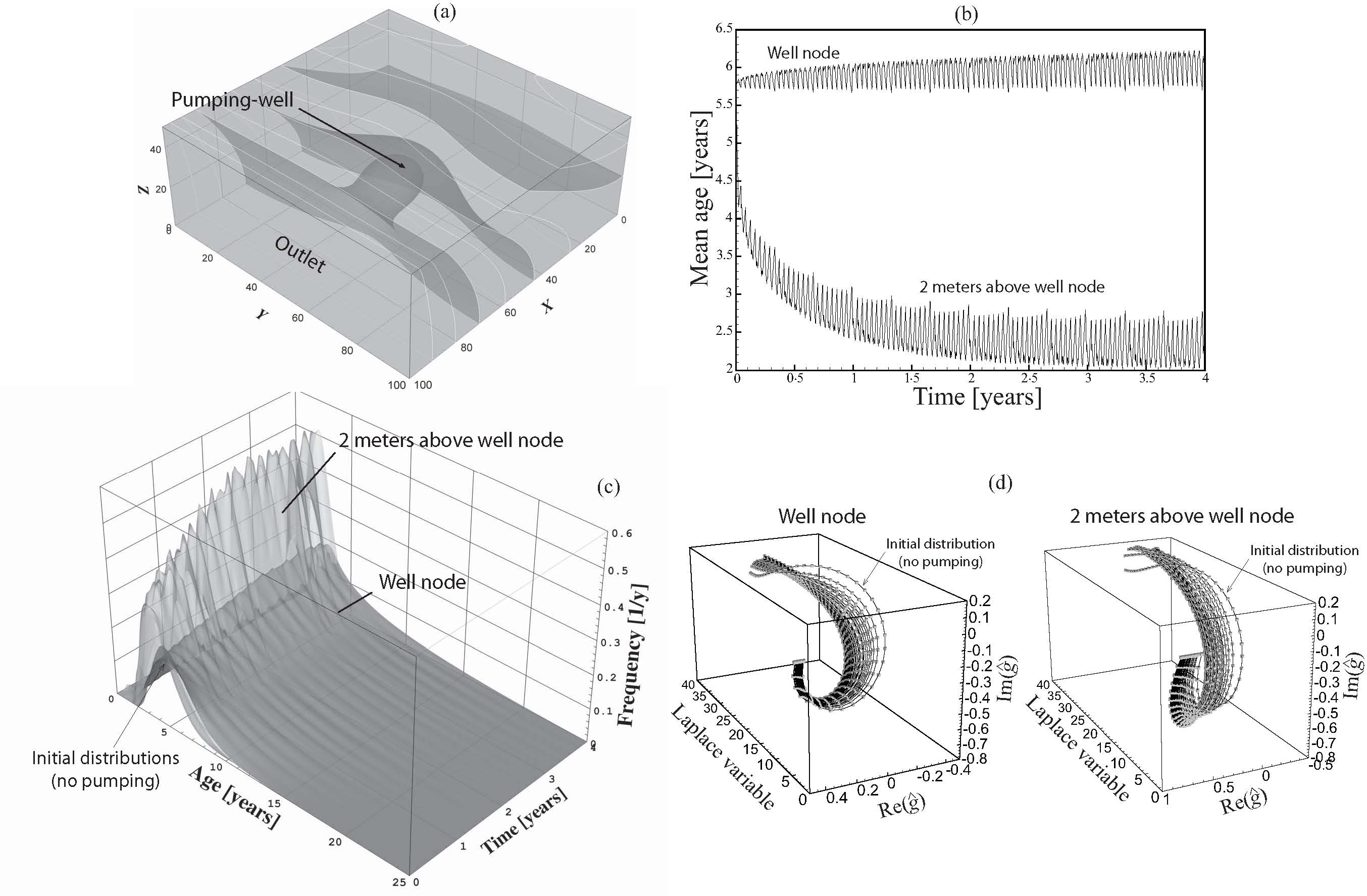}
\caption{Temporal evolution of mean age and of the age frequency distribution at the pumping-well. (a) Hydraulic solution at a given simulation time illustrating the effect of the pumping; (b) Mean age evolution; (c) Age distribution evolution; (d) Age distribution evolution in the complex Laplace space.\label{fig:fig9}}
\end{figure*}

Our last example corresponds to the situation of the activation of a pumping-well inside a homogeneous aquiferous block of 100 by 100 meters and of thickness 50 meters. The initial state is obtained by recharging the aquifer at steady-state at a uniform infiltration rate $2 \times 10^{-8}$~m/s on top of the model. The outlet is modelled using a fixed hydraulic head of 50 meters prescribed along one vertical face, as indicated in Figure~\ref{fig:fig9}a. As opposed to the previous 2-D example, we now make use of a natural, undisturbed initial condition for the transient solution. The mesh is made of 8-noded brick elements of size $1 \times 1 \times 1$ meter. Hydraulic conductivity is set to $K = 10^{-5}$~m/s, storage coefficient is $S = 3 \times 10^{-5}$~1/m, porosity is 0.1, longitudinal dispersivity is $\alpha_L = 1$~m, transverse horizontal dispersivity is $\alpha_{T,h} = 0.1$~m, transverse vertical dispersivity is $\alpha_{T,v} = 0.01$~m, and the coefficient of molecular diffusion is $D_m = 2.3 \times 10^{-9}~\textrm{m}/\textrm{s}^2$.\protect\\
The transient-state situation starts with the activation of a pumping-well located at $\{x, y, z\} = \{50, 50, 25\}$, using an extraction function that globally corresponds to cycles of pumping during 1 week at a rate ranging between $2 \times 10^{-5}~\textrm{m}^3/\textrm{s}$ and $1.5 \times 10^{-4}~\textrm{m}^3/\textrm{s}$, and no-pumping during 1 week. The remaining boundary conditions are kept unchanged. Figure~\ref{fig:fig9}a shows the hydraulic solution at a given simulation time, using three isosurfaces of hydraulic head. Figure~\ref{fig:fig9}b shows the mean age solution. Mean age tends to meet pseudo-equilibrium (i.e. an average situation considering the pump rate fluctuations) after 3 to 4 years. The solution displays important variations in the neighborhood of the well. At the well location, mean age slightly increases due to the capture of water molecules that have traveled from upstream locations and that enrich the mixing process with older ages. The small increase (from 5.75 years at initial time to around 6 years) at the well may be explained by the result of a mixing that equilibrates a component of newly captured older ages and a component of newly captured young ages coming from top locations above the well. The temporal variations of mean age due to pump rate variations are of about 0.5 year. Only 2 meters above the well location, mean age strongly decreases (from 5.45 years at initial time to around 2.5 years, i.e. more than 50\% or relative change) due to the capture of a major component of young water that infiltrates above the well. The temporal variations of mean age are of about 0.75 year. These important changes in mean age point out the risk associated to the interpretation of apparent ages deduced from isotopic methods, particularly when interpretation is carried out for estimating flow velocities or simply constraining the calibration of a flow model. \protect\\
The age distribution solution makes use of $N_L = 2 n + 1 = 41$ discrete Laplace variables and of the implicit time-integration scheme with a constant stepsize of 1 day. The effect of activating the pump on the age distribution recorded at the well location is shown in Figure~\ref{fig:fig9}c. The shape of the distributions confirms the interpretation based on the analysis of mean age solution. The distribution at the well node is more dispersed than the distribution recorded 2 meters above, and displays a bigger contribution of young water. The pump fluctuations are well-marked and revealed by peak variations.

%
\section{Conclusions and Outlooks}\label{sec:conclusions}
A novel and promising numerical scheme for solving the water age distribution problem in environmental transient flow domains of arbitrary complexity has been elaborated. The presented scheme can be of high interest for various interconnected fields like atmospheric and ocean circulation modelling or surface and subsurface hydrology. The main characteristic of the algorithm is that a reduction in problem dimension is obtained by mathematical transformation of the age dimension using the Laplace transform operator. A standard time-marching procedure is then applied to the Laplace transformed age distributions to simulate their temporal evolution. The main advantage of transforming the age distribution function is to be relieved from the need of discretization in the age-axis. The age dimension is therefore compressed in the complex Laplace space, which is independent on the age magnitude. A direct consequence is that, as opposed to the few existing methods, age distributions that display very big age values (e.g. of the order of 1 Million years or more, as observed in deep sedimentary basins) can be modelled without any specific limitation. The proposed algorithm is consequently particularly well-suited to treat hydrologic or atmospheric problems involving predictions of circulations and mass transfers over very large time-scales (in both the clock-time and age coordinates). Moreover, the algorithm allows for the integration of nonlinearities affecting a flow solution, such as the ones induced by variably saturated conditions in subsurface hydrology, surface-subsurface coupling of flow dynamics, or buoyancy effects encountered in thermohaline processes, and can also be applied to compressible homogeneous fluid flow problems. The approach can be implemented using the classical numerical integration schemes like the finite element, finite volume or finite differences methods.\protect\\
Since the method combines a Laplace transform of the age equation in the age-coordinate to a classical finite-difference approximation of the time evolution in the clock-time coordinate, the well-known numerical artifacts generated by badly controlled time-marching schemes and affecting the standard advection-dispersion transport equations also need to be considered. Consequently, the constraints on the space- and time-discretization have to be considered to avoid numerical instabilities (like the constraints on the P\'{e}clet and Courant numbers).\protect\\
The approach computational demand can be compared to that required by the solution of a transport problem involving independent species with first-order decay (however without reactions between the species), the number of species being equal to the number of used Laplace variables (ideally between 10 and 20, at maximum 40). Thus in order to be of real practical interest for handling problems of big size (i.e. with meshes/grids of more than 1 to 10 Million unknowns) and high parameter contrasts, the need of parallelization techniques applied to the assembling and solving parts of the solution can be foreseen.\protect\\
The presented numerical experiments do not have the pretention of providing an exhaustive analysis of the effects of time-varying flow conditions on age distributions. They are rather used to validate and illustrate the proposed method. They however show that transient age distributions undergoing variations in recharge and/or discharge conditions can reveal to be very complex and that realistic initial conditions for age can be a critical aspect to be resolved. These experiments point out probable severe complications in the field of age dating with single and multiple-tracer techniques and flow model calibration based on isotopic interpretations of age. Theoretical analysis of the time-dependent behavior of age distributions should be carried out in order to gain insights on fluid and mass transfer processes over various time-scales and space-scales. \protect\\
%
\begin{acknowledgments}
I thank Timothy Ginn and one anonymous reviewer for their helpful and constructive commentaries. I also acknowledge the French National Agency for Radioactive Waste Management (Andra) for having financially supported this research.
\end{acknowledgments}
%

%
\appendix
\section*{Appendix: Derivation of a 1-D Pseudo-Analytical Solution for Transient Age} \label{ap:A}
We consider a semi-infinite domain $(0\leq x < \infty)$ with uniform fluid velocity $v$. At any position $x$, we suppose that the initial age distribution is of Gaussian type, e.g. as a result of an equilibrated distribution in a uniform and constant velocity field $u$:
\begin{equation}\label{eq:Aeq0}
g_u(x,\tau) = \frac{x}{\sqrt{4 \pi \bar D} \tau^{3/2}} \exp \left ( -\frac{( x - u \tau )^2}{4 \bar D \tau} \right )
\end{equation}
where $\bar D = \alpha_L u + D_m$. The transient-state age distribution $g = g(x,t,\tau)$ is solution of the 1-D version of equation~(\ref{eq:5Dage1}):
\begin{equation}\label{eq:Aeq1}
\frac{\partial g}{\partial t} = - v \frac{\partial g}{\partial
x} + D \frac{\partial^2 g}{\partial x^2} - \frac{\partial g}{\partial \tau}
\end{equation}
in which the dispersion coefficient reads $D = \alpha_L v + D_m$, and with the initial and boundary conditions:
\begin{equation}\label{eq:Aeq1ic1}
g(x,0,\tau) = g_u(x,\tau)
\end{equation}
\begin{equation}\label{eq:Aeq1bc1}
g(0,t,\tau) = \delta(\tau)
\end{equation}
\begin{equation}\label{eq:Aeq1bc2}
- \left.D \frac{\partial g}{\partial x} \right|_{x = \infty } = 0
\end{equation}
To obtain a solution to that problem, we first apply a Laplace transform to equations~(\ref{eq:Aeq1}) to~(\ref{eq:Aeq1bc2}) in the $\tau$-dimension:
\begin{subequations}\label{eq:Aeq2}
\begin{equation}\label{eq:Aeq21}
\frac{\partial \hat g}{\partial t} = - v \frac{\partial \hat g}{\partial
x} + D \frac{\partial^2 \hat g}{\partial x^2} - s \hat g + g(x,t,0)
\end{equation}
\begin{equation}\label{eq:Aeq22}
\hat g(x,0,s) = \hat g_u(x,s) = \exp \left ( \frac{x u}{2 \bar D} \left ( 1 - \sqrt{1 + \frac{4 \bar D s}{u^2}} \right ) \right )
\end{equation}
\begin{equation}\label{eq:Aeq23}
\hat g(0,t,s) = 1
\end{equation}
\begin{equation}\label{eq:Aeq24}
- \left.D \frac{\partial \hat g}{\partial x} \right|_{x = \infty } = 0
\end{equation}
\end{subequations}
where $s$ is the complex Laplace variable dual to age $\tau$, and where $\hat g = \hat g(x,t,s)$ is the transformed state of $g(x,t,\tau)$. Now applying a second Laplace transform to equations~(\ref{eq:Aeq21}),~(\ref{eq:Aeq23}) and~(\ref{eq:Aeq24}) in the $t$-dimension yields:
\begin{subequations}\label{eq:Aeq4}
\begin{equation}\label{eq:Aeq41}
- v \frac{\partial \tilde g}{\partial
x} + D \frac{\partial^2 \tilde g}{\partial x^2} -\left ( r + s \right ) \tilde g + \hat g_u = 0
\end{equation}
\begin{equation}\label{eq:Aeq4bc13}
\tilde g(0,r,s) = \frac{1}{r}
\end{equation}
\begin{equation}\label{eq:Aeq4bc23}
- \left.D \frac{\partial \tilde g}{\partial x} \right|_{x = \infty } = 0
\end{equation}
\end{subequations}
where $r$ is the complex Laplace variable dual to $t$, $\tilde g = \tilde g(x,r,s)$ is the transformed state of $\hat g(x,t,s)$, and where we assumed that $g(x,t,0) = 0$. This assumption is valid since for this specific configuration of a flow system, the frequency for having zero ages is not nil only at $x = 0$ where it is Dirac. The solution to this boundary value problem reads:
\begin{equation}\label{eq:Aeq5}
\tilde g(x,r,s) = \frac{r e^{ \frac{x (1 - \beta_u(s))}{2 \alpha_L}
} + s \omega e^{ \frac{x (1 - \beta_v(r+s))}{2 \alpha_L} }}{r (r + s
\omega)}
\end{equation}
with
\begin{subequations}\label{eq:Aeq51}
\begin{equation}\label{eq:AeqOmega}
\omega = 1 - \frac{v}{u}
\end{equation}
\begin{equation}\label{eq:Aeqbu}
\beta_u(z) = \sqrt{1+\frac{4 \alpha_L z}{u}}
\end{equation}
\begin{equation}\label{eq:Aeqbv}
\beta_v(z) = \sqrt{1+\frac{4 \alpha_L z}{v}}
\end{equation}
\end{subequations}
and where molecular diffusion has been neglected ($D_m = 0$) for a sake of simplicity. The coefficient $w$ represents the relative change in velocity. Inversion of equation~(\ref{eq:Aeq5}), $\hat g(x,t,s) = \textrm{L}^{-1} \{\tilde
g(x,r,s), r, t \}$, yields:
\begin{equation}\label{eq:Aeq61}
\hat g(x,t,s) = \hat \gamma_0(x,t,s)
+ \hat \gamma_1(x,t,s) - \hat \gamma_2(x,t,s)
\end{equation}
with
\begin{subequations}\label{eq:Aeq62}
\begin{align}\label{eq:Aeq621}
\hat \gamma_0(x,t,s) = {} & e^{ \frac{x(1-\beta_u(s))}{2 \alpha_L} - s \omega t}
\end{align}
\begin{align}\label{eq:Aeq622}
\hat \gamma_1(x,t,s) = {} & \frac{1}{2} e^{
\frac{x(1-\beta_v(s))}{2 \alpha_L}} \textrm{erfc} \left ( \frac{x -
\beta_v(s) v t}{2 \sqrt{\alpha_L v t}} \right ) \\
 + & \frac{1}{2} e^{\frac{x(1+\beta_v(s))}{2 \alpha_L}} \textrm{erfc} \left ( \frac{x +
\beta_v(s) v t}{2 \sqrt{\alpha_L v t}} \right ) \nonumber
\end{align}
\begin{align}\label{eq:Aeq623}
\hat \gamma_2(x,t,s) = {} & \frac{1}{2} \hat \gamma_0(x,t,s) \textrm{erfc} \left (
\frac{x - \beta_u(s) v t}{2 \sqrt{\alpha_L v t}} \right ) \\
 + & \frac{1}{2} e^{\frac{x(1+\beta_u(s))}{2 \alpha_L} - s \omega t} \textrm{erfc} \left (
\frac{x + \beta_u(s) v t}{2 \sqrt{\alpha_L v t}} \right ) \nonumber
\end{align}
\end{subequations}
One can verify that the mass in the system is always 1 by evaluating the zero-order age moment of $g(x, t, \tau)$, $\int_0^\infty {g(x, t, \tau) d\tau} = \hat g(x,t,s \rightarrow 0) = 1$. The solution~(\ref{eq:Aeq61}) can be expressed in dimensionless form by
letting $L$ be a characteristic length, and by using the change of variables $X = \frac{x}{L}$, $T = \frac{t}{\tau_v} = \frac{v t}{L}$, $Pe = \frac{v L}{D} = \frac{L}{\alpha_L}$, and $z = \tau_v s$. Note that $\tau_v =
\frac{L}{v}$ denotes the mean turnover time (or reservoir mean residence time) of the new
equilibrated flow system (under the flow regime associated to velocity $v$), and that $Pe$ is the P\'{e}clet number. The dimensionless solution becomes:
\begin{equation}\label{eq:Aeq6ad}
\hat g(X,T,z) = \hat \gamma_0(X,T,z) +
\hat \gamma_1(X,T,z) - \hat \gamma_2(X,T,z)
\end{equation}
with
\begin{subequations}\label{eq:Aeq6ad1}
\begin{align}\label{eq:Aeq6ad11}
\hat \gamma_0(X,T,z) = {} & e^{ \frac{X Pe(1-\beta_w(z))}{2} - z \omega T}
\end{align}
\begin{align}\label{eq:Aeq6ad12}
\hat \gamma_1(X,T,z) = {} & \frac{1}{2} e^{ \frac{X
Pe(1-\beta_0(z))}{2}} \textrm{erfc} \left ( \frac{X -
\beta_0(z) T}{2 \sqrt{{T \mathord{\left/
 {\vphantom {T {Pe}}} \right.
 \kern-\nulldelimiterspace} {Pe}}}} \right ) \\
 & + \frac{1}{2} e^{ \frac{X Pe(1+\beta_0(z))}{2}
} \textrm{erfc} \left ( \frac{X + \beta_0(z)
T}{2 \sqrt{{T \mathord{\left/
 {\vphantom {T {Pe}}} \right.
 \kern-\nulldelimiterspace} {Pe}}}} \right ) \nonumber
\end{align}
\begin{align}\label{eq:Aeq6ad13}
\hat \gamma_2(X,T,z) = {} & \frac{1}{2} \hat \gamma_0(X,T,z) \textrm{erfc} \left ( \frac{X -
\beta_w(z) T}{2 \sqrt{{T \mathord{\left/
 {\vphantom {T {Pe}}} \right.
 \kern-\nulldelimiterspace} {Pe}}}} \right ) \\
 & + \frac{1}{2} e^{ \frac{X
Pe(1+\beta_w(z))}{2} - z \omega T} \textrm{erfc} \left ( \frac{X + \beta_w(z)
T}{2 \sqrt{{T \mathord{\left/
 {\vphantom {T {Pe}}} \right.
 \kern-\nulldelimiterspace} {Pe}}}} \right ) \nonumber
\end{align}
\end{subequations}
and we have $\hat g(x,t,s) = \tau_v \hat g(X,T,z)$. When time tends toward infinity, the new equilibrated age distribution
$g_v(X,\zeta) = g(X,\infty,\zeta) = \textrm{L}^{-1} \{ \hat g(X,\infty,z), z, \zeta \}$ is found:
\begin{equation}\label{eq:Aeqgv}
g_v(X,\zeta) =
\frac{X \sqrt{Pe}}{\sqrt{4 \pi} \zeta^{3/2}} \exp \left ( -\frac{Pe ( X
- \zeta )^2}{4 \zeta} \right )
\end{equation}
with the dimensionless age dimension variable $\zeta = \frac{\tau}{\tau_v}$. The first-order age moment of equation~(\ref{eq:Aeq61}) gives the transient-state mean age $a(x,t)$:
\begin{align}\label{eq:Aeqma1}
a(x,t) = {} & \lim_{s=0} \left ( - \frac{\partial \hat g(x,t,s)}{\partial s} \right ) \\
& = \frac{x}{u} + w t \nonumber \\
& + \frac{w}{2 v} (x - v t) \textrm{erfc} \left (
\frac{x - v t}{2 \sqrt{\alpha_L v t}} \right ) \nonumber \\
& - \frac{w}{2 v} (x + v t) e^{ \frac{x}{\alpha_L} } \textrm{erfc} \left (
\frac{x + v t}{2 \sqrt{\alpha_L v t}} \right ) \nonumber
\end{align}
which satisfies the properties $a(x,0) = \frac{x}{u}$ and
$a(x,\infty) = \frac{x}{v}$, and that is solution of the 1-D form of the mean age equation (see equation~(\ref{eq:MA ADE})), $\frac{\partial a}{\partial t} = - v \frac{\partial a}{\partial
x} + D \frac{\partial^2 a}{\partial x^2} + 1$, when the boundary condition $a(0,t) = 0$ is used.
Dimensionless mean age finally reads:
\begin{align}\label{eq:Aeqmaadim}
a(X,T) = {} & ( 1 - w ) X + w T \\
& + \frac{w}{2} (X-T)  \textrm{erfc} \left ( \frac{X -
T}{2 \sqrt{{T \mathord{\left/
 {\vphantom {T {Pe}}} \right.
 \kern-\nulldelimiterspace} {Pe}}}} \right ) \nonumber \\
 & - \frac{w}{2}(X+T) e^{ X Pe } \textrm{erfc} \left ( \frac{X +
T}{2 \sqrt{{T \mathord{\left/
 {\vphantom {T {Pe}}} \right.
 \kern-\nulldelimiterspace} {Pe}}}} \right ) \nonumber
\end{align}

\end{article}

\end{document}